\documentclass[acmsmall,screen]{acmart}

\usepackage{subfigure}

\usepackage{multirow}

\usepackage{hyperref}
\usepackage{makecell}
\usepackage{algorithm}
\usepackage{algorithmic}
\usepackage{amsmath}

\usepackage{lettrine}
\usepackage{tcolorbox}
\tcbuselibrary{breakable} 

\usepackage{xcolor}
\usepackage{soul}  

\usepackage{pifont}

\usepackage[pagewise]{lineno}

\AtBeginDocument{%
  }

\newcommand{\TGen}{LLMCFG-TGen}

\setcopyright{acmlicensed}
\copyrightyear{2026}
\acmYear{2026}
\acmDOI{0000000.0000000}

\acmJournal{JACM}
\acmVolume{77}
\acmNumber{7}
\acmArticle{777}
\acmMonth{06}

\newcounter{RubingCommentCounter}
\newcounter{ZhenzhenCommentCounter}
\newcounter{DaveCommentCounter}
\newcommand{\ddu}[1]{
    \stepcounter{DaveCommentCounter}
    \textcolor{blue}{\textit{/**Dave's comment [\arabic{DaveCommentCounter}]: I don't understand the intended meaning in the next sentence. Please revise/delete/explain. **/}}
}

\newcommand{\dns}[1]{
    \stepcounter{DaveCommentCounter}
    \textcolor{blue}{\textit{/**Dave's comment [\arabic{DaveCommentCounter}]: I'm not sure that I have captured the intended meaning in the next sentence. Please check/confirm. **/}}
}

\begin{document}

\title[LLMCFG-TGen]{LLMCFG-TGen: Using LLM-Generated Control Flow Graphs to Automatically Create Test Cases from Use Cases}

\author{Zhenzhen Yang}
\email{3240002362@student.must.edu.mo}
\orcid{0009-0009-9358-7675}
\affiliation{
  \institution{School of Computer Science and Engineering, Macau University of Science and Technology}
  \city{Macao SAR}
  \postcode{999078}
  \country{China}  
}
\affiliation{
  \institution{School of Artificial Intelligence, Zhejiang Polytechnic University of Mechanical and Electrical Engineering}
  \city{Hangzhou}
  \state{Zhejiang}
  \postcode{310053}
  \country{China}
}

\author{Chenhui Cui}
\email{3230002105@student.must.edu.mo}
\orcid{0009-0004-8746-316X}
\affiliation{
  \institution{School of Computer Science and Engineering, Macau University of Science and Technology}
  \city{Macao SAR}
  \postcode{999078}
  \country{China}
}

\author{Tao Li}
\email{abelli@buaa.edu.cn}
\orcid{0009-0001-7413-9692}
\affiliation{
  \institution{School of Artificial Intelligence (Institute of Artificial Intelligence), Beihang University}
  \city{Beijing}
  \postcode{100191}
  \country{China}
}

\author{Rubing Huang}
\email{rbhuang@must.edu.mo}
\orcid{0000-0002-1769-6126}
\affiliation{
  \institution{School of Computer Science and Engineering, Macau University of Science and Technology}
  \city{Macao SAR}
  \postcode{999078}
  \country{China}
}
\affiliation{
  \institution{Macau University of Science and Technology Zhuhai MUST Science and Technology Research Institute}
  \city{Zhuhai}
  \state{Guangdong}
  \postcode{519099}
  \country{China}  
}

\author{Nan Niu}
\email{nan.niu@unf.edu}
\orcid{0000-0001-5566-2368}
\affiliation{
  \institution{School of Computing, University of North Florida}
  \city{Jacksonville}
  \state{FL}
  \country{USA}
  \postcode{32224}
}

\author{Dave Towey}
\email{dave.towey@nottingham.edu.cn}
\orcid{0000-0003-0877-4353}
\affiliation{
  \institution{School of Computer Science, University of Nottingham Ningbo China}
  \city{Ningbo}
  \state{Zhejiang}
  \postcode{315100}
  \country{China}
}

\author{Shikai Guo}
\email{shikai.guo@dlmu.edu.cn}
\orcid{0000-0002-8554-6365}
\affiliation{
  \institution{School of Information Science and Technology, Dalian Maritime University}
  \city{Dalian}
  \state{Liaoning}
  \postcode{116026}
   \country{China}   
}

\renewcommand{\shortauthors}{Yang et al.}

\begin{abstract}
    Appropriate test-case generation is critical in software testing and significantly impacts testing quality.
    \textit{Requirements-Based Test Generation} (RBTG) derives test cases from software requirements to verify whether or not the system's behavior aligns with user needs and expectations. 
    Requirements are often documented in \textit{Natural Language} (NL), with use-case descriptions being a popular method for capturing functional behaviors and interaction flows in a structured, readable form.
    Recently, \textit{Large Language Models} (LLMs) have shown strong potential for automating test generation directly from NL requirements (including use-case descriptions).
    However, current LLM-based approaches may not provide comprehensive, non-redundant coverage.
    They may also fail to capture complex conditional logic in requirements, resulting in incomplete test cases.
    To address these limitations, we propose a new end-to-end approach that automatically generates test cases from NL use-case descriptions, called 
    \textit{Test Generation based on LLM-generated Control Flow Graphs} (\TGen\/).
    \TGen\/ comprises the following three main steps: 
    (1) CFG Generation: 
    An LLM transforms a use case into a structured, JSON-based \textit{Control Flow Graph} (CFG) that encapsulates all potential branches;
    (2) Test-Path Extraction: 
    The generated CFG is traversed to extract a finite set of execution paths; and
    (3) Test-Case Creation: 
    Finally, the execution paths are used to generate the test cases.
    To evaluate the proposed approach, we conducted experiments on six use-case datasets covering diverse application domains. 
    The results show that LLMs can effectively construct well-structured CFGs from NL use cases.  
    Compared with two baseline approaches, \TGen\/ generates more complete and structurally consistent test cases by more effectively capturing requirement behaviors and execution flows.
    Both LLM-based and practitioner-based evaluations further confirm that \TGen\/ produces more comprehensive and logically coherent test cases while reducing manual effort.
    The findings suggest that coupling LLM-based semantic reasoning with structured modeling effectively bridges the gap between NL requirements and systematic test generation.     
\end{abstract}



\received{8 June 2026}

\maketitle

\section{Introduction}

Software testing is an important part of the software development process, aimed at ensuring the quality and reliability of software systems. 
Test-case generation is one of the most challenging testing tasks, often accounting for 40\% to 70\% of the overall testing effort \cite{verma2013generation}.
Automated test-case generation has emerged as a promising approach to reducing both effort and cost \cite{jin2021generation}. 
Test cases can be derived from various sources, including requirements specifications \cite{yang2025requirements}, design documents \cite{zakeriyan2021towards}, source code \cite{pan2025hamster}, and other artifacts such as bug reports \cite{plein2024automatic}. 
Code-based test generation derives tests from the program structure, focusing on implementation-level coverage \cite{ammann2017introduction, pressman2005software}. 
In contrast, \textit{Requirements-Based Test Generation} (RBTG) 
derives test cases directly from requirements, ensuring alignment with the expected system behavior, without requiring access to the source code \cite{yang2025requirements}.
This paradigm is particularly relevant in the context of the widespread adoption of \textit{Test-Driven Development} (TDD) \cite{sisomboon2026automated}, where tests are expected to be defined early, and to reflect the intended system behavior.
In most approaches, RBTG produces high-level test cases that capture system functionality and behavior.
RBTG enables test design in the early stages of development, which not only facilitates customer validation, but also guides the system implementation.
By enabling early-stage test design, RBTG helps developers form a clearer understanding of the expected system behavior, reducing the risk of error propagation to later stages \cite{el2010developing}.
These high-level test cases differ from low-level test scripts in that they specify {\em what} should be tested, in terms of system functionality and behavior, rather than {\em how} it should be executed:
They serve as an intermediate behavioral specification that helps testers systematically derive detailed, executable test scripts.

Early RBTG approaches typically generated test cases from requirements expressed in formal or semi-formal notations.
Examples of this include \textit{Unified Modeling Language} (UML)  diagrams \cite{mustafa2021automated}, and \textit{Finite State Machines} (FSMs) \cite{liu2010new}.
Model-based approaches, in which abstract models are constructed from requirements and then used to derive test cases, have since emerged as the most popular strategy \cite{yang2025requirements}. 
However, model-based strategies have not been widely adopted in industry because developing and maintaining models can be complex and costly \cite{lim2024test, yang2025requirements, ahmad2019model}.
Due to its simplicity, \textit{Natural Language} (NL) remains popular for documenting requirements in industry \cite{dwarakanath2012litmus}.
However, NL can be ambiguous, inconsistent, and incomplete, making automated test generation difficult.
Use-case descriptions can help to address this:
They follow standardized templates \cite{alrawashed2019automated}, and capture essential behavioral details through \textit{main}, \textit{alternative}, and \textit{exception} flows \cite{jiang2011automation, binder2000testing}.
Use cases are also widely adopted in practice to validate and refine business processes with stakeholders, thereby improving clarity and reducing ambiguity \cite{wang2020automatic}.
This has made textual use cases a popular choice for automated test-case generation \cite{yang2025requirements}.
Most methods involving use cases use an intermediate representation, such as an \textit{Extended Finite State Machine} (EFSM) \cite{jiang2011automation} or 
\textit{Use-Case Maps} (UCMs) \cite{kesserwan2023transforming}. 
However, these models can be difficult to build and maintain, and require that use cases be sufficiently detailed and precisely specified to support effective test automation \cite{wang2020automatic}.

Due to their strong NL understanding and reasoning abilities, \textit{Large Language Models} (LLMs) have become increasingly popular in software engineering, supporting tasks such as requirements analysis, code generation, test-case generation, and defect prediction \cite{hou2024large}. 
LLMs can generate test cases directly from NL requirements \cite{wang2024software}, using zero-shot \cite{bhatia2024system, li2025evaluating} and one-shot prompting \cite{hasan2025automatic}, or through fine-tuning \cite{kang2024automated, xue2024llm4fin}.
However, extracting specific information from complex requirements documents can lead to hallucinations \cite{rahman2024automated}. 
It is also difficult to (reliably) generate large numbers of test cases \cite{kirinuki2024chatgpt}, with the process often producing redundant tests \cite{bhatia2024system}. 
Furthermore, fine-tuning approaches may perform well only in specific domains and lack generality \cite{xue2024llm4fin}. 
In summary, current LLM-based test-generation approaches suffer from key limitations.

To address these limitations, we propose an automated approach for generating test cases from NL use-case descriptions, \textit{Test Generation based on LLM-generated Control Flow Graphs} (\TGen\/).
A \textit{Control Flow Graph} (CFG) is used as an intermediate representation to enable systematic enumeration of requirements-level execution paths.
\TGen\/ is driven by two carefully designed prompts, and comprises three main steps:
\begin{itemize}
    \item [(1)] 
    CFG Generation: 
    An LLM transforms the use case into a structured CFG, represented in JSON format, capturing all potential behavioral branches. 
    The generated CFG is validated to ensure structural correctness and completeness before further processing. 
    This step is guided by Prompt \#1, which provides instructions and an explicit algorithm for CFG construction.
    
    \item [(2)] 
    Test-Path Extraction: 
    The system then traverses the constructed CFG to extract a set of valid execution paths.
    A \textit{Depth-First Search} (DFS) algorithm identifies distinct node-level paths, and each node identifier in a path is replaced with its corresponding NL statement.

    \item [(3)] 
    Test-Case Creation: 
    The statement-level paths are converted into detailed test cases by the LLM. 
    Each test case is formatted according to predefined templates and then presented to the user for review and confirmation.
    This step is guided by Prompt \#2, specifying both the instructions and the test‑case structure.
\end{itemize}
Without restricting use-case expressiveness or requiring manual intervention, \TGen\/ supports single-click test-generation through a web interface, producing comprehensive, non-redundant, and highly readable test cases.

To evaluate the proposed approach, we conducted a study using six multi-domain datasets covering 60 use cases. 
We compared the generated CFGs with manually-constructed reference CFGs, assessing their structural similarity and alignment. 
Compared with two baseline approaches, \TGen\/ generated more complete and less redundant test cases, better capturing the required behaviors and execution flows. 
An LLM-based assessment and a practitioner study further demonstrated that \TGen\/ produces more consistent and comprehensive test cases, while also reducing manual effort.

This paper makes the following contributions:
\begin{itemize}
    \item 
    We propose \TGen\/, a CFG-guided approach that generates test cases from NL use-case specifications through structured control-flow modeling and path-based extraction, without requiring predefined templates for use-case specifications.
    
    \item 
    We introduce an LLM-based behavior-alignment assessment method for abstract test cases that quantifies the alignment between generated test cases and reference execution paths across multiple behavioral dimensions.

    \item 
    We report on an empirical study to assess the effectiveness and reliability of the proposed approach:
    This study involved a quantitative analysis, a practitioner evaluation with statistical significance analysis, and a cost-effectiveness analysis across different LLMs.
    
    \item 
    We provide a web-based prototype of \TGen\/ to support reproducibility and facilitate experimental evaluation.
\end{itemize}

The rest of this paper is structured as follows:
Section~\ref{sec:backgroundandrelatedwork} introduces the background to the proposed approach, and reviews related work.  
Section~\ref{sec:approach} provides a detailed description of the methodology.  
Section~\ref{sec:experimentalsetup} outlines the experimental setup.  
The results of the experiments and the threats to validity are examined in Section~\ref{sec:experimentalresults}.  
Section~\ref{sec:discussion} discusses the strengths and weaknesses of the approach.
Finally, Section~\ref{sec:conclusion} concludes the paper and identifies some potential future work.

\section{Background and Related Work
\label{sec:backgroundandrelatedwork}}

This section introduces some background information for the study, including use cases, CFGs, and types of test cases.
Some related work is also discussed.

\subsection{Use-Case Description}
Use cases can document many elements of the software development process, including business workflows \cite{cockburn2008writing}. 
They can facilitate requirements discussions, help define functional specifications, and record system design decisions~\cite{cockburn2008writing,kulak2012use}. This paper focuses on their use to describe functional requirements \cite{dwarakanath2012litmus}.

A use-case description is a structured textual artifact that describes the system’s behavior in response to interactions initiated by external actors \cite{wang2020automatic, craig2002systematic}.
It can capture possible sequences of events that lead to achieving a specific goal \cite{cockburn2008writing}.
Compared to full requirements specifications, use-case descriptions are shorter narratives that focus on a single function.
They differ from UML use-case diagrams, which graphically depict the relationships among actors and use cases (or among related use cases) \cite{cockburn2008writing}.
They have a strict syntax, but lack behavioral detail, making them suitable for early-stage communication \cite{kulak2012use}.
When used for testing, use-case descriptions provide much (though not all) of the information needed to develop a complete system test suite \cite{binder2000testing}. 

Writing high-quality use cases can be challenging, because no precise rules exist for their composition, and a diverse range of templates and styles have been produced \cite{de2023echo}. 
\citet{jacobson1987object} used a paragraph-style format with fields such as name, actors, preconditions, postconditions, basic flow, and alternate flows. 
Later work proposed constraints and authoring guidelines: 
Cockburn's template, for example, has been widely adopted and is recognized as one of the most comprehensive formats \cite{cockburn2008writing,tiwari2020use}. 
Our proposed approach is template-agnostic and can accommodate use cases written in diverse formats, without requiring any structural standardization.

\subsection{Control Flow Graphs}
Graphical models are widely used in requirements engineering to support communication among stakeholders, and to guide system design~\cite{ferrari2024model}.
A CFG is an abstract representation of all possible execution paths through a component or system~\cite{board2014standard}. 
It is typically defined as a directed graph \( G = (S, E) \), where \( S \) denotes the set of nodes and \( E \subseteq S \times S \) represents the set of directed edges~\cite{lim2020approach}.
Nodes typically represent basic blocks, and edges denote control transfers between them.

In our context, the nodes correspond to the main, alternative, and exception steps within a use case, while the edges represent control transitions between steps (including sequencing, branching, and looping). 
Edges may be annotated with guard conditions that reflect the decision logic.
(It is assumed that the CFG has no unreachable nodes or edges.) 
An execution path is any path from a unique root node to one or more exit nodes \cite{lafi2021automated}:
They serve as the foundation for our path-based test-case creation.

\subsection{Types of Test Cases}

\begin{figure}
    \centering
    \includegraphics[width=0.9\linewidth]{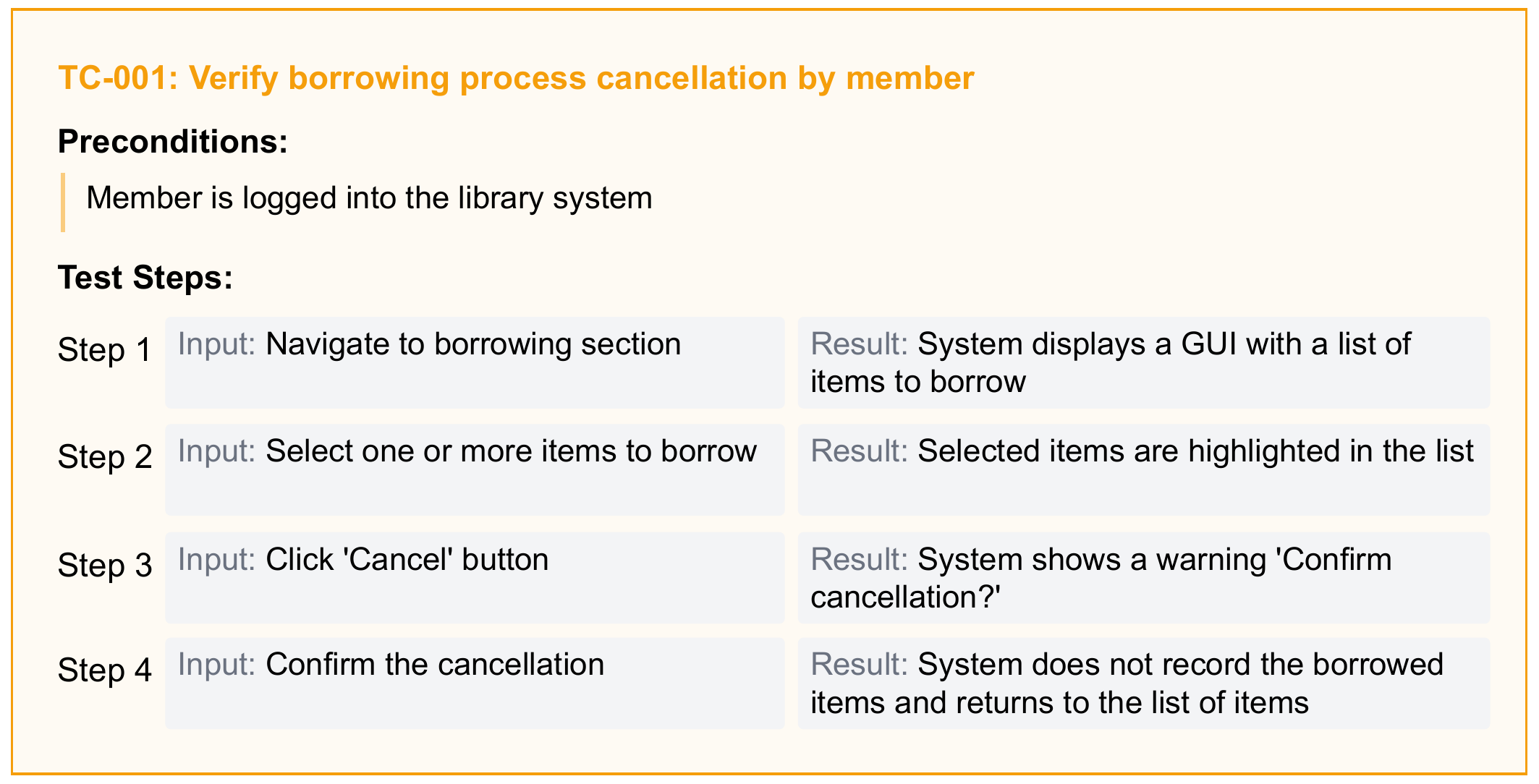}
    \caption{Example test case.}
    \Description{Example test case.}
    \label{fig:testcasetemplate}
\end{figure}

A \textit{test case} defines a specific condition to be evaluated, and consists of the inputs, conditions of execution, and expected outcomes \cite{board2014standard, craig2002systematic}.
The expected outcomes specify the results used to evaluate test execution \cite{barr2014oracle, ammann2017introduction}.
Test cases can be categorized into three types \cite{yang2025requirements}:
\textit{Abstract test cases} capture test logic at a high level, without specifying concrete (implementation-level) input values or expected outcomes;
\textit{concrete test cases} specify detailed, executable inputs, along with the corresponding expected results;
and \textit{test scenarios} describe functional behaviors, or business processes, to be validated from a user's perspective.
Our study used textual \textit{abstract test cases}, which typically include a title, precondition, test steps, and expected results \cite{yin2024leveraging}.
Figure~\ref{fig:testcasetemplate} presents an example of a test case created by our approach.

\subsection{Related Work}
Traditional testing approaches derive test cases from use-case descriptions using intermediate models. 
\citet{frohlich2000automated} outlined a strategy that maps use-case elements to UML state machines for deriving test suites, but did not provide an implementation or case study.
\citet{de2006generating} proposed using a formal CSP (\textit{Communicating Sequential Processes}) behavioral model, built from use cases, as a basis for test generation. 
However, this remained a conceptual proposal, without implementation or empirical validation.
\citet{alrawashed2019automated} defined a pipeline that refines Cockburn-style use cases and converts them into a CFG for test generation.
Transformations have also targeted UML activity diagrams \cite{gutierrez2008case,tiwari2015approach}, meta-models \cite{zhang2014systematic,allala2019towards,allala2022generating}, and \textit{Use Case Test Models} (UCTMs) \cite{wang2020automatic, chenail2025test}:
These approaches typically use predefined use-case templates or require manual refinement of use-case descriptions before transforming them into intermediate models. 
The intermediate models themselves were manually constructed or heavily guided by human input, adding complexity and effort to the process. 
Abstract test cases could then be derived from the manually-constructed models.

Recent studies have shown the effectiveness of LLMs for generating tests directly from NL requirements \cite{wang2024software}:
They can do this using zero-shot \cite{bhatia2024system, li2025evaluating} and one-shot prompting \cite{hasan2025automatic}, or through fine-tuning \cite{kang2024automated, xue2024llm4fin}.
However, these approaches may often suffer from issues of completeness, redundancy, and clarity
---
due to hallucinations, and the inherent difficulty of extracting accurate information from lengthy or ambiguous requirements \cite{rahman2024automated}. 
\citet{mathur2023automated} incorporated use-case context and conversational cues to enhance LLM prompting. 
\citet{chenail2025test} investigated LLM-based and hybrid approaches for generating test cases from use-case specifications in IoT systems:
They found that hybrid strategies combining domain-specific design with LLM capabilities achieved the best performance, highlighting the benefits of integrating LLM reasoning into structured generation pipelines.
\citet{de2024agora} introduced the AGORA pipeline, leveraging LLMs to automate the generation of acceptance tests from use-case descriptions:
They could achieve results comparable to those of manual testing by experienced engineers. 
Overall, LLM-based methods can reduce the dependency on test-design expertise, minimize manual effort, and improve efficiency:
These advantages support scalable test-generation for complex systems.
Nevertheless, there are still challenges to achieving sufficient coverage and to eliminating redundancy \cite{bhatia2024system}.

To address these challenges, we combine the reasoning capability of LLMs with the structural rigor of CFGs as an intermediate representation. 
Unlike previous work \cite{alrawashed2019automated}, our approach imposes no constraints on the format of use cases and requires no manual intervention. 
Integrating CFG-guided generation into the LLM pipeline enhances coverage, reduces redundancy, and produces more consistent and interpretable test cases. 
The next section presents the proposed approach in detail, describing its architecture and key processing steps.

\section{Approach
\label{sec:approach}}
This section presents our proposed approach, \TGen\/.

\subsection{Architecture Overview}

\begin{figure}
  \centering
  \includegraphics[width=\linewidth]{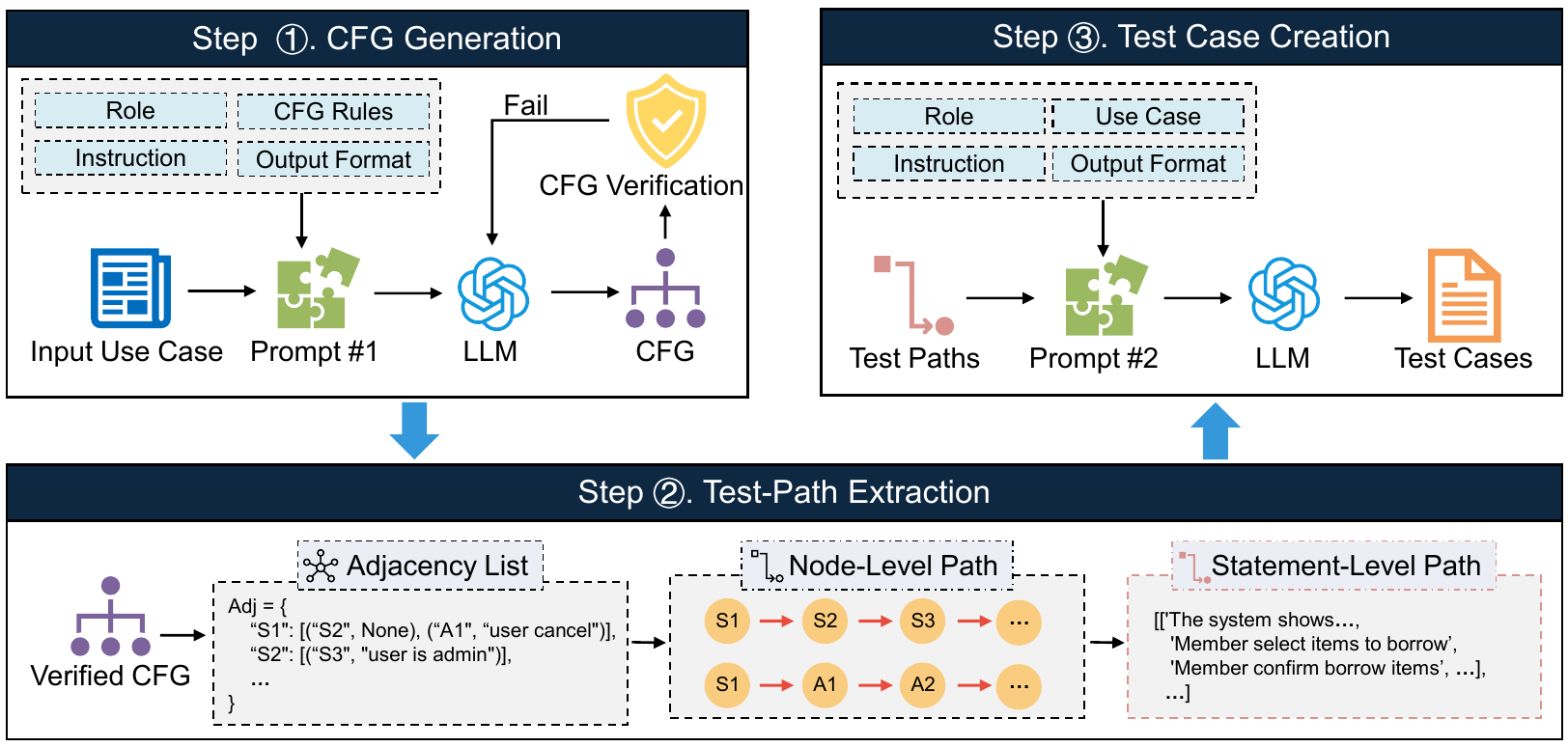}
  \caption{\TGen\/ architecture overview.}
  \Description{Overview of the proposed approach.}
  \label{fig:Fig3_1_approach}
\end{figure}

Figure \ref{fig:Fig3_1_approach} shows the framework of \TGen\/, which, without any preprocessing or format constraints, transforms an NL use-case description into a set of high‑coverage test cases.
The process comprises the following three main steps:

\textbf{Step \ding{182} CFG Generation}: 
The NL use-case description is passed to an LLM, which, guided by a prompt that integrates the description with predefined control‑flow rules, produces a CFG (in JSON format).
The CFG is then checked for completeness, with the LLM being re-prompted until a fully-connected graph is obtained.

\textbf{Step \ding{183} Test-Path Extraction}: 
Starting from the root node, the CFG is traversed to enumerate execution paths. 
To prevent infinite expansion when facing cycles, the traversal is bounded by pruning paths whose nodes are revisited.

\begin{figure}
  \centering
  \includegraphics[width=\linewidth]{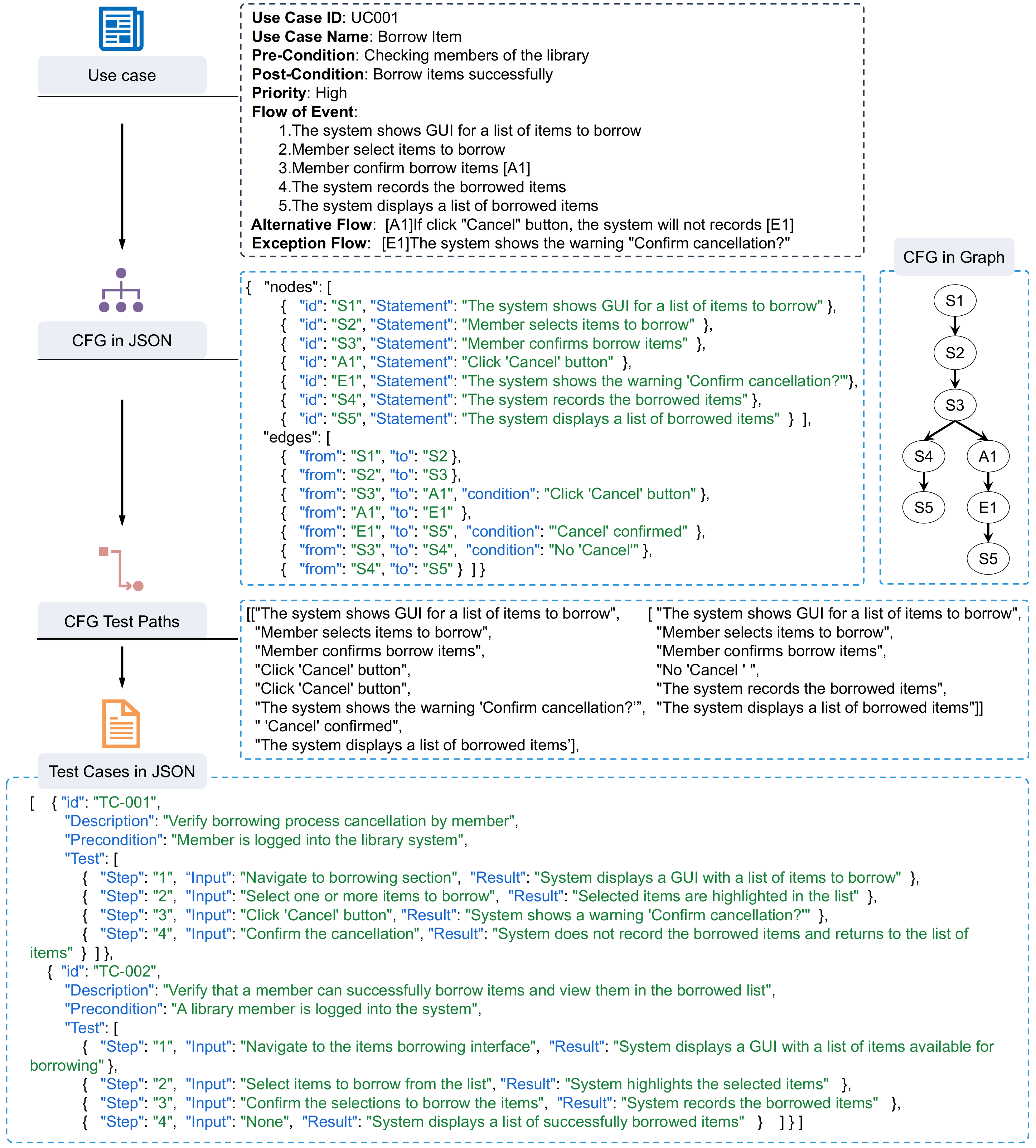}
  \caption{End-to-end process, from an NL use case, to the LLM-generated CFG and derived test cases.}
  \Description{A pipeline showing the step-by-step transformation from a NL use case into abstract test cases.}
  \label{fig:Fig3_3_usecase_cfg}
\end{figure}

\textbf{Step \ding{184} Test Case Creation}: 
The LLM converts each extracted path into a detailed test case, comprising a title, preconditions, and an ordered sequence of step-and-expected-result pairs.
Figure \ref{fig:Fig3_3_usecase_cfg} shows an example use case, with its generated CFG, corresponding test paths, and structured test cases (formatted in JSON).
We have also developed a user-friendly web application that automates the workflow, enabling users (without expertise in LLMs or software testing) to efficiently generate high-quality test cases.

\subsection{Step \ding{182}: CFG Generation}
\textbf{Step \ding{182}} consists of three tasks: 
Prompt construction; 
CFG generation using an LLM; and 
CFG verification.

\subsubsection{Prompt Construction}
\label{sec:prompt}

Prompt engineering \cite{wang2024software} involves interacting with the LLM, designing a task description that is passed as input to the LLM.
Zero-shot and few-shot prompting are among the most widely used strategies \cite{arora2024generating}.
With zero-shot prompting, the model receives only the task description (without any examples) and must generate the output based only on these instructions. 
Few-shot prompting, in contrast, provides the model with a small number of examples, which help to guide it towards the expected response or output.

Our study used a streamlined zero-shot prompting strategy, involving two prompts: 
Prompt \#1 to generate the CFG; and
Prompt \#2 to create the corresponding test cases.
Prompt \#1 provides unambiguous guidance to the LLM by combining a concise, well-structured instruction set with an explicit CFG-construction algorithm.
It also instructs the LLM to present the generated CFG in JSON format.

\begin{figure}
  \centering
  \includegraphics[width=0.9\linewidth]{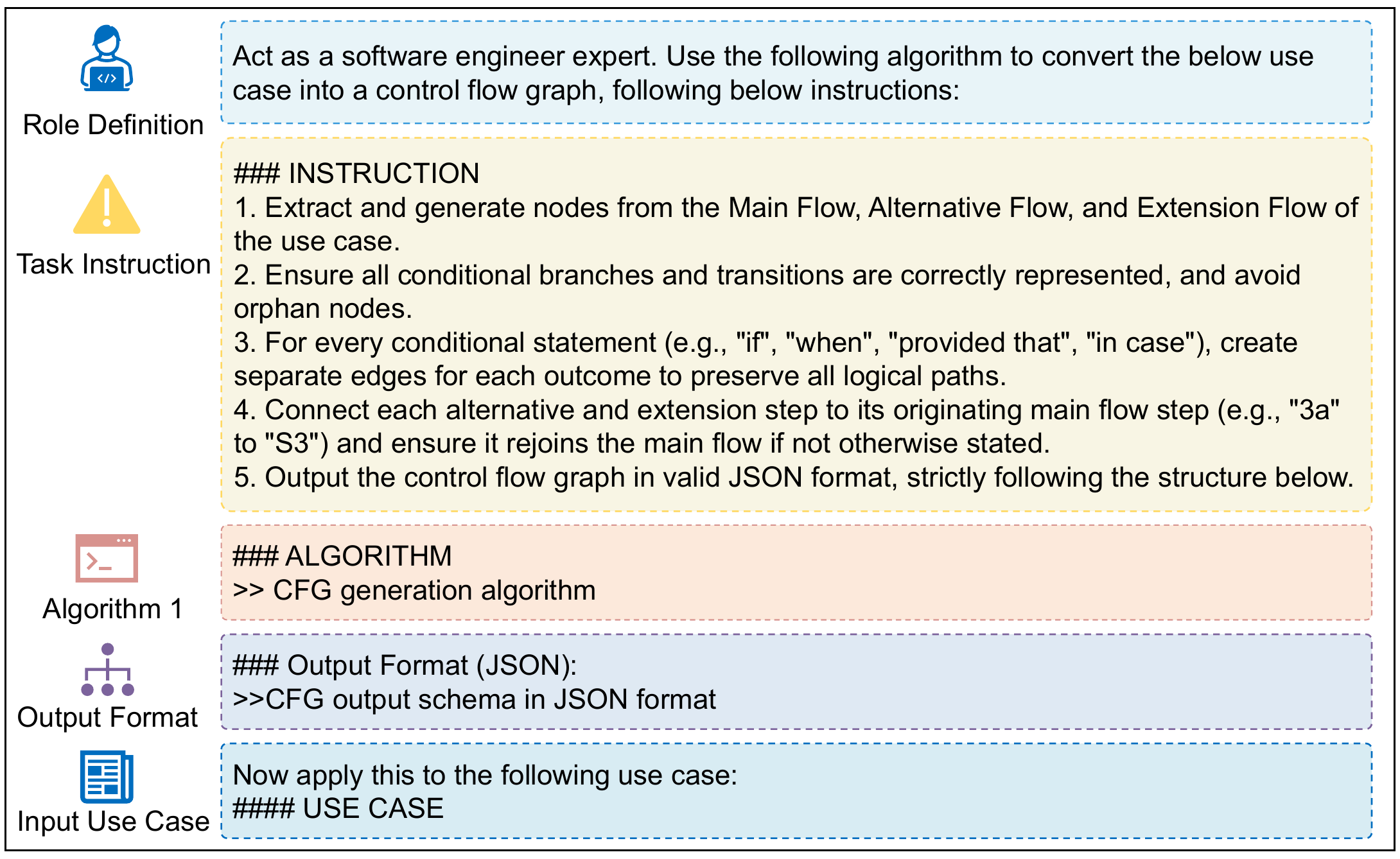}
  \caption{LLM prompt for generating the CFG from a use case.}
  \Description{LLM prompt for generating the CFG from a use case.}
  \label{fig:Fig3_2_prompt}
\end{figure}

Figure~\ref{fig:Fig3_2_prompt} shows how Prompt \#1 is used to generate a CFG from an NL use-case description. 
The prompt consists of five components, the last of which is the input use case, with the other four being:
\begin{enumerate}
    \item 
    \textit{Role definition}: 
    This specifies that the LLM should act as a software engineering expert.
    
    \item 
    \textit{Task instructions}: 
    These provide high-level operational guidance.
    They describe how to extract nodes from the main, alternative, and extension flows;
    how to handle conditional branches; and
    how to ensure that there are no orphan nodes \cite{jaffar2010path}.
    
    \item 
    \textit{Algorithm specification} (Algorithm~\ref{alg:cfg}): 
    The algorithm provides the LLM with a formal mapping process that translates the extracted elements into CFG nodes and edges.
    It includes detailed rules for conditional branching and CFG edge construction.
    
    \item 
    \textit{Output specification}: 
    This requires that the CFG be returned in valid JSON, following a predefined schema with \texttt{nodes} and \texttt{edges}.    
\end{enumerate}

\begin{algorithm}
\caption{CFG Generation}
\label{alg:cfg}
\begin{algorithmic}[1]
\renewcommand{\algorithmicrequire}{\textbf{Input:}}
\renewcommand{\algorithmicensure}{\textbf{Output:}}
\renewcommand{\algorithmiccomment}[1]{\hfill $\triangleright$ #1}
\REQUIRE Steps in main flow and other flows of use-case description
\ENSURE CFG $G=(S,E)$ with root $R$
\STATE Preprocess: Merge all steps into one ordered sequence
\STATE Let $S=\{S_1,S_2,\ldots,S_n\}$ be the resulting ordered nodes
\STATE Set root $R := S_1$; Initialize edge sequence $E := \varnothing$
\STATE Add edge $E_1 := S_1 \to S_2$
\FOR{each step $S_i$}
    \IF{$S_{i+1}$ contains a condition}
        \STATE add edge $E_i := S_i \to S_{i+1}$ \COMMENT{For the true condition}
        \STATE add edge $E_{i+1} := S_i \to S_{i+2}$ \COMMENT{For the false condition}
    \ELSE
        \STATE add edge $E_i := S_i \to S_{i+1}$
    \ENDIF
\ENDFOR
\end{algorithmic}
\end{algorithm}

Algorithm~\ref{alg:cfg} formalizes the mapping process from the use case to its corresponding CFG. 
Nodes represent steps in the use case, and edges are established based on sequential and conditional relationships. 
Alternative and extension flows are explicitly linked to the 
main flow. 
This ensures the generation of a complete and valid CFG, enabling the direct parsing and seamless integration into all subsequent processing stages.

\subsubsection{CFG Generation Using an LLM}
CFGs are generated using an LLM (such as \textit{GPT-5.4} \cite{openai_gpt54_2026}). 
The proposed \TGen\/ approach is model-agnostic, meaning that it can be adapted to other LLMs without additional modifications.
A comparative evaluation of multiple LLMs is presented in Section \ref{sec:RQ4}.
Several key parameters affect the LLM's behavior: 
The \textit{temperature} \cite{de2024agora} (ranging from 0 to 1) regulates output randomness, with higher values encouraging creative variation, and lower values leading to more deterministic outputs; 
while the \textit{top-$p$} parameter \cite{de2024agora} (also ranging from 0 to 1) limits token choices to the smallest set whose cumulative probability surpasses $p$, filtering out unlikely tokens.
Following OpenAI’s guidelines \cite{openai_api_reference}, and previous work \cite{de2024agora}, we set the temperature to $0.0$;
top-$p$ to its default value of $1.0$; and 
both frequency and presence penalties to their default values of $0.0$.
This aimed to ensure deterministic and reproducible generation.

The prompt (Section \ref{sec:prompt}) containing a use case is input to the LLM, which then generates the CFG. 
Figure \ref{fig:Fig3_3_usecase_cfg} shows an example use case, with its generated CFG, corresponding test paths, and structured test cases formatted in JSON format.
The JSON-encoded CFG consists of \textit{nodes} and \textit{edges}.
Conditional nodes have at least two child nodes, each labeled with a statement corresponding to a step in the use-case scenario.
When applicable, edges can also include \textit{conditions} extracted from the logical flow of the use case.

\subsubsection{CFG Verification}
Although the LLM is explicitly instructed to avoid producing isolated nodes or edges, its inherent non-determinism \cite{ouyang2025empirical, atil2025non} may still result in structurally invalid outputs. 
A post-generation validation step is therefore applied.

Each CFG generated from a use case has a single designated root node.
A CFG is structurally invalid if any of the following conditions hold:
\begin{enumerate}
    \item 
    {\textit{Isolated node}}: 
    A node does not appear in any edge.
    
    \item 
    {\textit{Unreachable node}}: 
    A node is not reachable from the designated root through graph traversal (e.g., DFS).
    
    \item 
    {\textit{Orphaned reference}}: 
    A node referenced in the \texttt{from} or \texttt{to} field of an edge does not exist in the node set.
\end{enumerate}

If a CFG is deemed invalid, the LLM is prompted to regenerate CFGs until a valid structure is obtained. 
The validated CFG is then used to extract the test paths.

\begin{algorithm}
\caption{Test-Path Extraction}
\label{alg:testpath}
\begin{algorithmic}[1]
\renewcommand{\algorithmicrequire}{\textbf{Input:}}
\renewcommand{\algorithmicensure}{\textbf{Output:}}
\renewcommand{\algorithmiccomment}[1]{\hfill $\triangleright$ #1}
\REQUIRE Verified CFG $G = (S, E)$, where each node $s \in S$ has an identifier and statement; each edge $(u, v) \in E$ may have an optional condition
\ENSURE Set of test paths $T$, each as an ordered list of textual steps
\STATE Let $root :=$ first node in $S$
\STATE Build adjacency list $Adj$ from $E$, storing $(neighbor, condition)$ for each edge
\STATE Initialize empty set $P := \varnothing$ to store all paths

\STATE Define recursive procedure \textsc{DFS}($curr, path$)
\IF{$curr \in path$}
    \STATE Append copy of $path$ to $P$ \COMMENT{Save current path when cycle detected to avoid infinite loops}
    \STATE \textbf{return}
\ENDIF
\STATE Append $curr$ to $path$  \COMMENT{Add current node to path}
\IF{$Adj[curr]$ is empty}
    \STATE Append $path$ to $P$ \COMMENT{Save complete path when path completion}
    \STATE \textbf{return}
\ENDIF

\FORALL{$(nbr, cond) \in Adj[curr]$} 
    \STATE $path' :=$ copy of $path$
    \IF{$cond \neq \texttt{null}$}
        \STATE Append \texttt{"condition: cond"} to $path'$  \COMMENT{Add condition step}
    \ENDIF
    \STATE \textsc{DFS}($nbr, path'$)   \COMMENT{Recursive exploration}
\ENDFOR
\STATE Call \textsc{DFS}($root, []$)    \COMMENT{Start traversal from root}
\STATE Map each node ID in $P$ to its corresponding statement
\STATE Insert each preserved condition as a separate step
\STATE Collect the resulting step sequences into $T$
\RETURN $T$
\end{algorithmic}
\end{algorithm}

\subsection{Step \ding{183}: Test-Path Extraction}
\begin{figure}
  \centering
  \includegraphics[width=\linewidth]{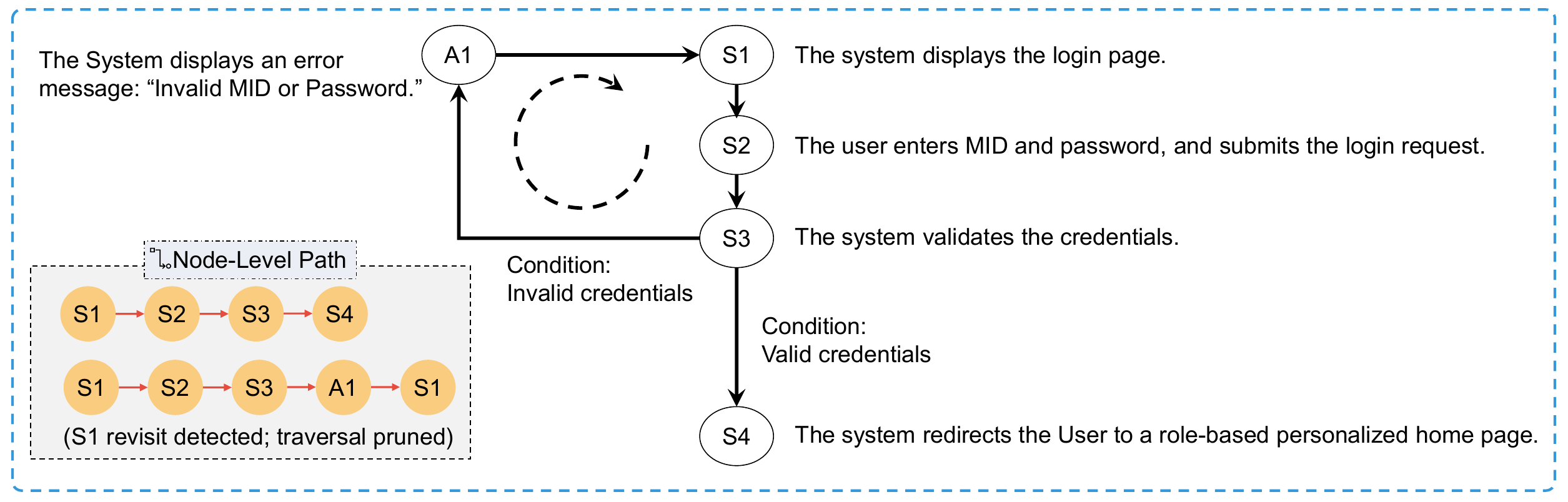}
  \caption{Example of test-path extraction with bounded DFS traversal and cycle pruning.}
  \Description{A diagram showing two extracted node-level paths from a CFG.}
  \label{fig:step2_path}
\end{figure}

Given a verified CFG $G =(S, E)$, this step derives a finite set of execution paths that preserve structural coverage over nodes and edges. 
A test path is defined as a sequence of nodes that starts at the entry node and either terminates at a leaf node, or upon detecting a cycle. 
To ensure finite CFGs, a path is terminated if a node reappears within the same execution trace. 
We construct a bounded set of execution paths through cycle truncation, consistent with the notion of simple paths in graph-based testing~\cite{ammann2017introduction}. 
As illustrated in Figure~\ref{fig:step2_path}, cyclic structures are represented up to the first repetition point, capturing loop-related behavior without redundant path expansion. 
This preserves cyclic behavior while preventing infinite path expansion:
Each bounded path corresponds to a complete executable scenario up to the first repetition point.

Algorithm~\ref{alg:testpath} formalizes the extraction procedure, which consists of three steps:
\begin{enumerate}
    \item 
    \textit{Adjacency-list construction}: 
    The edge set $E$ is transformed into an adjacency list $Adj$, where each source node is mapped to its successor nodes along with optional transition conditions.
    
    \item 
    \textit{Path enumeration}: 
    A DFS is performed over $Adj$ starting from the root node. 
    During traversal, a set of visited nodes is maintained, to detect revisiting. 
    A path is terminated and recorded when a node is revisited or when a terminal node is reached. 
    This ensures finite enumeration while preserving both forward and cyclic transitions.
    
    \item 
    \textit{Path translation}: 
    Each node identifier in the extracted paths is replaced with its corresponding NL statement.
    Edge conditions are preserved as explicit steps, producing readable execution scenarios for subsequent test-case generation.
\end{enumerate}

\subsection{Step \ding{184}: Test Case Creation
\label{sec:Step3}}

\begin{figure}
  \centering
  \includegraphics[width=0.8\linewidth]{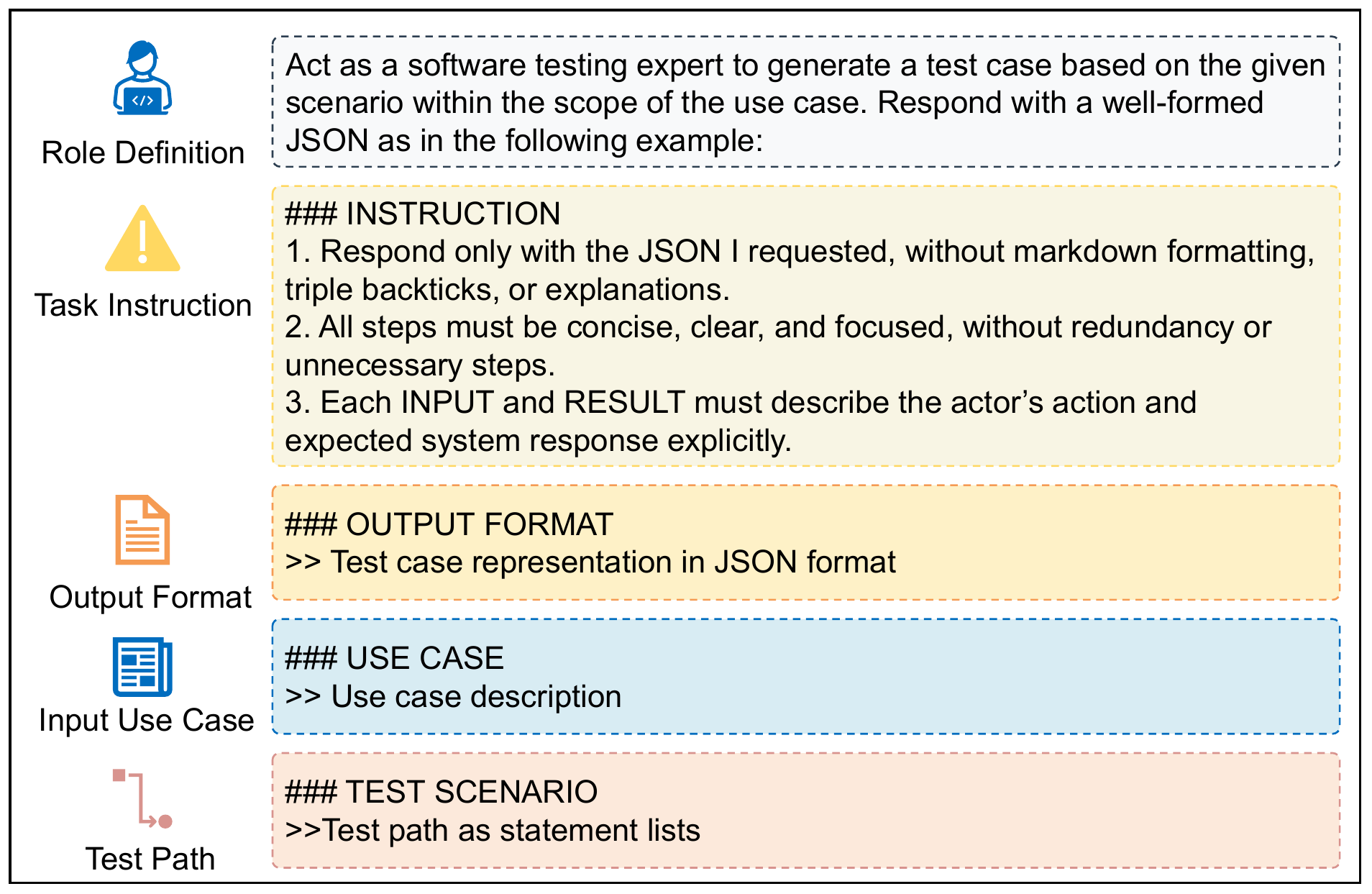}
  \caption{LLM prompt for generating test cases based on a test path.}
  \Description{LLM prompt for generating test cases based on a test path.}
  \label{fig:Fig3_2_prompt2}
\end{figure}

The test-creation step involves passing each textual test path (from Step~\ding{183}) and the original use-case description to the LLM. 
Figure~\ref{fig:Fig3_2_prompt2} shows the prompt (Prompt \#2), which defines the role, specifies the instructions, and includes a test-case example. 
The LLM produces a complete test case for each test path. 
Test inputs may contain ``Empty'' or ``None'' entries, indicating that no explicit input value is required for that step.
Any LLM output not conforming to the expected format cannot be rendered correctly in the web interface:
In such cases, the LLM is prompted to regenerate the test case, ensuring strict adherence to the output format, and enabling a fully automated workflow.

This per-path strategy ensures that the created test cases are internally coherent, directly traceable to a specific execution path in the CFG, and provide comprehensive coverage of the original use case.
This design enhances clarity, preserves contextual integrity, and ensures semantic completeness.

\section{Experimental Setup
\label{sec:experimentalsetup}}

\subsection{Research Questions
\label{section:RQs}}
Four research questions (RQs) guided the evaluation of the proposed approach's effectiveness. 
These RQs examined the correctness of the LLM-generated CFGs; the coverage and quality of the resulting test cases; and the impact of using different LLMs.  
\begin{itemize}
    \item 
    \textbf{RQ1:} 
    How accurate are the LLM-generated CFGs compared to a manually-constructed ground truth?
    This RQ evaluates whether or not the generated CFGs capture the intended control flow defined by the use case.   
    
    \item 
    \textbf{RQ2:} 
    How effectively do test cases generated by \TGen\/ align with the ground-truth execution structure, compared with the baseline methods?
    This RQ examines how well the generated test cases cover the intended behavior space defined by the ground-truth CFG:
    It considers both the quantity and quality, in terms of coverage completeness and redundancy.
    
    \item 
    \textbf{RQ3:} 
    How does the quality of \TGen\/ test cases compare with the baseline methods?
    This RQ focuses on the perceived usefulness and clarity of the test cases, as assessed through human evaluation.
    
    \item 
    \textbf{RQ4:} 
    How do different LLMs compare in terms of cost-effectiveness?
    This RQ evaluates the cost-effectiveness of different LLMs, with cost measured by token consumption and execution time across the end-to-end generation pipeline.
\end{itemize}

\subsection{Datasets}

\begin{table}
    \centering
    \renewcommand{\arraystretch}{1.25}
    
    \caption{Use-Case Datasets Utilized in the Experiment}
    \label{tab:datasets}
    \footnotesize
    \begin{tabular}{l p{3cm} p{4.8cm} c c}
        \hline
        \textbf{Dataset} & \textbf{Source} & \textbf{Description} & \textbf{\#UCs} & \textbf{\#Words (Total/Range)}\\
        \hline
        UC1 & \textit{AGORA: An Approach for Generating Acceptance Test Cases from Use Cases}~\cite{de2024agora} & Remote monitoring of patients using wearable devices (e.g., smartwatch) that transmit physiological data to the system. & 12 & 1782 (111 -- 218) \\[3pt]
        UC2 & Use cases collected from multiple prior studies~\cite{gutierrez2008case, jiang2011automation, intana2023approach, kesserwan2023transforming, allala2022generating, intana2020syntest, tiwari2015approach, wang2020automatic} & Case studies from different domains such as \textit{search}, \textit{seller submits an offer}, etc.; overall well-structured and representative. & 13 & 1793 (57 -- 287) \\[3pt]
        UC3 & \textit{ECHO: An Approach to Enhance Use Case Quality Exploiting Large Language Models}~\cite{de2023echo} & Patient measurement management system; use cases refined and standardized by domain engineers. & 5 & 1202 (176 -- 304) \\[3pt]
        UC4 & \textit{PURE: A Dataset of Public Requirements Documents}~\cite{ferrari2017pure} & From the requirements document \texttt{triangle.pdf}, describing core functions of a simple game (e.g., startup, save/load, AI setting). & 12 & 1324 (85 -- 183) \\
        UC5 & \textit{PURE: A Dataset of Public Requirements Documents}~\cite{ferrari2017pure} & From the requirements document \texttt{003-pnnl.pdf} (Automated Diagnostics Software Requirements Specification); preprocessed use case. & 6 & 9428 (1042 -- 2506) \\
        UC6 & \textit{iTrust Medical Records System Requirements}~\cite{massey2012legal} & Derived from the iTrust system; preprocessed use case. & 12 & 14,827 (690 -- 1882) \\
        \hline
        \textbf{\textit{Sum}} & -- & -- & \textbf{60} & \textbf{30,356 (57 -- 2506)} \\
        \hline
    \end{tabular}
\end{table}

To answer the RQs, we conducted experiments using multiple use-case datasets. 
Six datasets were collected from previous studies and public repositories:
The key details are summarized in Table~\ref{tab:datasets}, where the datasets are referred to as UC1--UC6:
\begin{itemize}
    \item 
    \textbf{UC1:} 
    UC1 has 12 use cases focusing on remote patient monitoring in the healthcare domain. 
    It includes non-human ``Device'' actors, emphasizing device-patient-system interactions, but lacks explicit ``Title'' or ``Postconditions'' fields (unlike UC3 and UC4). 
    
    \item 
    \textbf{UC2:} 
    UC2 has 13 use cases that cover cross-domain scenarios (including e-commerce, logistics, and automotive).
    Unlike the single-domain focus of UC1, UC3, and UC4, UC2 is flexible, with domain-specific elements (e.g., ``Bounded Alternative Flow'' for automotive use cases):
    This helps test the adaptability of the proposed method.
    
    \item 
    \textbf{UC3:} 
    UC3 comprises five use cases, and is a manually-refined healthcare dataset focused on vital-sign management.
    It focuses on human-human interactions (in contrast to UC1’s device-centric design), and has the most standardized format (each use case includes ``ID'', ``Title'', and detailed ``Error Handling'' elements).
    
    \item 
    \textbf{UC4:} 
    UC4 has 12 use cases targeting the game domain, including gamer-oriented operations (e.g., adding new AI) and high-level abstract use cases.
    The UC4 format integrates user roles into functional descriptions and includes unique ``Comments'' for cross-referencing (a feature absent from the other datasets).

    \item 
    \textbf{UC5:} 
    UC5 comprises six use cases derived from the PURE dataset~\cite{ferrari2017pure}. 
    It describes an automated diagnostics system, including administrative operations, such as starting and stopping diagnostic processing.
    The dataset was preprocessed to remove noise, resolve ambiguous statements, and decompose overly long use cases into smaller units with a single, well-defined functionality.

    \item 
    \textbf{UC6:} 
    UC6 includes 12 use cases derived from the iTrust Medical Records System~\cite{massey2012legal}. 
    It covers patient records, access control, and clinical data handling. 
    The dataset was preprocessed to remove noise, clarify ambiguous statements, and decompose overly complex use cases (to ensure that each use case captures a single, well-defined functionality).
\end{itemize} 

These datasets span diverse domains and vary in specification detail, ensuring sufficient representativeness for evaluating the CFG generation, test-path extraction, and test-case creation. 

\subsection{Baseline Approaches
\label{sec:baseline}}
Two baseline approaches were used in our evaluations:
(1) Direct test-case creation from use cases using LLMs, denoted LLM (Direct); and 
(2) AGORA \cite{de2024agora}:
\begin{itemize}
    \item 
    \textbf{LLM (Direct):} 
    Using LLMs to create test cases directly is a common approach \cite{hasan2025automatic, li2025evaluating}.
    This involves providing the LLM with a prompt containing the requirements and asking it to generate test cases that adhere to the specified format.
    Our evaluation prompt had four main components: 
    A role definition; 
    instructions for creating the test cases;
    an output format; and 
    the original use case. 
    The design of this prompt aligns with \TGen\/ Prompt \#2 (Section \ref{sec:Step3}).
    The prompt was provided to the LLM, and its test-case generation capability was assessed.
    
    \item 
    \textbf{AGORA:} 
    AGORA uses LLMs in a two-step pipeline to automatically generate acceptance test cases \cite{de2024agora}.
    In the first step, AGORA analyzes the use case to identify testable flows (main, alternative, and exception) and produces a list of test cases.
    In the second step, it generates detailed information for each test-case entry (including preconditions, participating actors, and step-by-step instructions with inputs and expected results).
\end{itemize} 

LLM (Direct) is a straightforward, commonly-used approach:
It enables the assessment of the core LLM capabilities for generating test cases from minimal input.
In contrast, AGORA is a structured and systematic methodology that ensures comprehensive test coverage through a multi-step pipeline. 
The combination of these two baselines enables a balanced and diverse assessment of other test-case creation techniques.

\subsection{Evaluation Procedure and Metrics}

\subsubsection{Overview}

\begin{figure}
    \centering
    \includegraphics[width=0.95\linewidth]{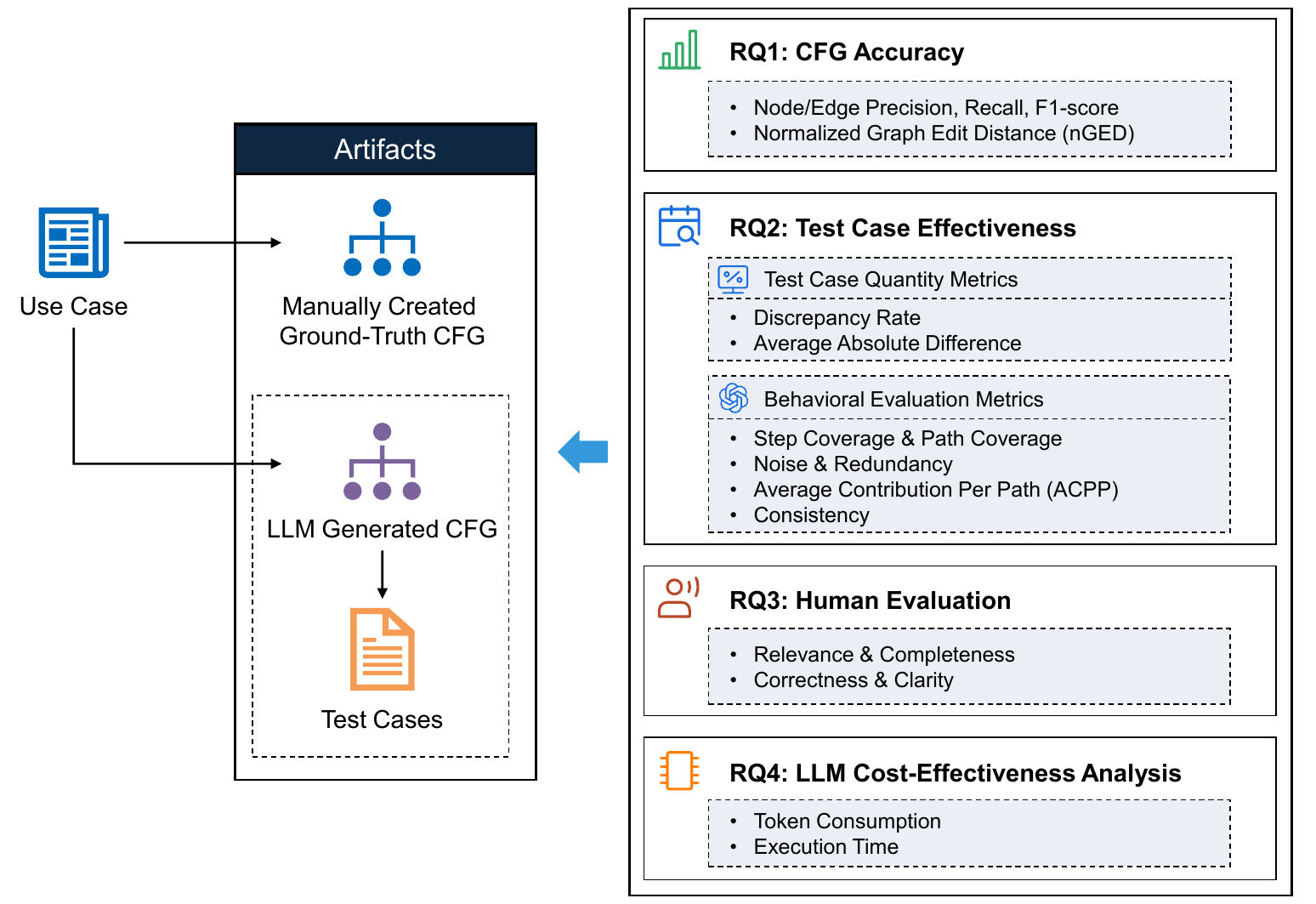}
    \caption{Overview of the evaluation framework.}
    \Description{The framework evaluates CFGs, test cases, human judgments, and LLM cost-effectiveness.}
    \label{fig:eval_framework}
\end{figure}

Figure~\ref{fig:eval_framework} shows our multi-level evaluation framework that assesses \TGen\/ from four 
complementary perspectives:
\begin{enumerate}
    \item 
    \textbf{Addressing RQ1 (CFG Accuracy):} We evaluate the similarity between LLM-generated CFGs and manually constructed ground-truth CFGs.
    
    \item 
    \textbf{Addressing RQ2 (Test Case Effectiveness):} We assess generated test cases in terms of quantity alignment and behavioral quality using quantitative metrics and behavioral evaluation. 
    
    \item 
    \textbf{Addressing RQ3 (Human Evaluation):} We evaluate the quality of generated test cases based on expert judgment.   

    \item 
    \textbf{Addressing RQ4 (LLM Cost–Effectiveness Analysis):} We examine trade-offs among LLMs under the same evaluation setting as RQ1 and RQ2. 
\end{enumerate}

This design allows each aspect of the approach to be evaluated using the most appropriate method.
Together, these perspectives provide a comprehensive assessment of the proposed approach.

The ground-truth CFGs were constructed independently by three doctoral students specializing in computer technology and applications. 
All three had previous industrial experience (two had worked in software testing and one in software development) and a solid understanding of use-case modeling and control-flow structures.  
Their expertise ensured the quality and consistency of the reference CFGs. 
Any discrepancy among their initial versions was resolved through collaborative discussion, resulting in a consolidated and reliable reference set for evaluation.

\subsubsection{CFG Accuracy Metrics (RQ1)}
\label{section:RQ1_eval}
We evaluated the accuracy of the LLM-generated CFGs by comparing them with manually-constructed ground-truth CFGs. 
The evaluation focused on the correspondence between generated and reference graphs, measured using two complementary metrics: 
\textit{Precision/Recall/F1-score} \cite{powers2020evaluation} and 
\textit{normalized Graph Edit Distance (nGED)} \cite{riesen2009approximate, abu2015exact}. 
These metrics evaluated CFG similarity from both node-level semantic matching and graph-level structural comparison.

\textbf{Precision/Recall/F1-score for Nodes and Edges:}
To assess node and edge correspondence, nodes in the generated CFG were aligned with those in the ground-truth CFG based on semantic similarity. 
The node similarity \cite{kirinuki2024chatgpt} was computed using cosine similarity over sentence embeddings produced by a pre-trained BERT-based model 
(\textit{all-mpnet-base-v2} \cite{huggingface_all_mpnet_base_v2}) within the \textit{SentenceTransformers} framework \cite{reimers2019sentencebert}. 
The Hungarian algorithm \cite{kuhn1955hungarian} (implemented using the \texttt{linear\_sum\_assignment} function in \texttt{SciPy} \cite{scipy2025}) was used to compute the optimal node alignment.
Two nodes were considered a match if their cosine similarity exceeded a threshold of $0.75$
---
this value was selected based on pilot experiments and previous work \cite{hasan2025automatic}.
Given the established node alignment, edge correspondence was evaluated by comparing the connectivity between matched node pairs.  
Let $TP$, $FP$, and $FN$ denote the numbers of true positives (correctly matched nodes/edges);
false positives (extra nodes/edges not in the ground truth); and 
false negatives (ground‑truth nodes/edges missing in the generated outputs), respectively.
The precision, recall, and F1-score were calculated as:
\begin{equation}
    Precision = \frac{TP}{TP + FP},
\end{equation}
\begin{equation}
    Recall = \frac{TP}{TP + FN},
\end{equation}
\begin{equation}
    F1\text{-}score = \frac{2 \times Precision \times Recall}{Precision + Recall}.
\end{equation}

Higher precision indicates that a larger proportion of generated nodes or edges are correctly matched to the ground-truth CFG, while higher recall indicates that fewer ground-truth elements are missed in the generated CFG. 
The F1-score balances precision and recall, reflecting overall matching quality.
Together, these metrics quantify the structural accuracy and completeness of the generated CFGs.

\textbf{Normalized Graph Edit Distance (nGED):}
Let $G_{gt}$ and $G_{gen}$ denote the ground-truth and generated CFGs, respectively.

To quantify structural differences between two CFGs, we adopted a normalized graph edit distance (nGED) measure \cite{riesen2009approximate, abu2015exact}. 
Instead of computing the absolute number of graph operations, nGED approximates the minimum cost of structural edits—including node/edge insertions, deletions, and substitutions—required to transform $G_{gen}$ into $G_{gt}$ under an optimal node alignment. 
To ensure comparability across CFGs of varying sizes, this edit cost is normalized by the total number of nodes and edges in both graphs, yielding a value between $0$ and $1$.

Given an optimal alignment between nodes in $G_{gen}$ and $G_{gt}$, we first map $G_{gen}$ into the node ID space of $G_{gt}$. 
Let $S_{gen}^*$ and $E_{gen}^*$ denote the aligned node and edge sets of the generated CFG.

We normalize this value using the total size of both graphs:
\begin{equation}
    nGED(G_{gt}, G_{gen}) =
    \frac{|S_{gt} \Delta S_{gen}^*| + |E_{gt} \Delta E_{gen}^*|}
    {(|S_{gt}| + |E_{gt}|) + (|S_{gen}| + |E_{gen}|)}.
\end{equation}

This normalization ensures that differences are measured relative to the combined structural complexity of both graphs.

The nGED values range between $0$ and $1$, where $0$ indicates a perfect structural match, and higher values indicate greater structural divergence.

\subsubsection{Test Case Effectiveness (RQ2)
\label{sec:metrics_rq2}}

Traditional evaluation of test cases often relies on manual coverage assessment, which is subjective and dependent on individual expertise~\cite{yin2024leveraging, lim2024test}. 
Some studies have also adopted lexical similarity metrics, such as BLEU and ROUGE~\cite{arora2024generating}: However, these metrics mainly capture surface-level text overlap, and are not sufficient for test-case evaluation~\cite{sulem2018bleu, li2024llms}.

Test cases represent behavioral specifications in which correctness depends on actions, conditions, and expected outcomes rather than on textual similarity. 
Consequently, semantically similar test cases may have low lexical overlap, while small textual differences may lead to substantially different behaviors~\cite{sulem2018bleu}.

To address these limitations, we evaluate test cases from two perspectives. 
Firstly, we assess the quantity by comparing the number of generated test cases with the number of ground-truth paths, using discrepancy-based metrics. 
Secondly, we evaluate coverage and quality through LLM-based behavioral assessment.

\textbf{Test Case Quantity Metrics.}
With a path-based testing strategy, each CFG path corresponds to a complete behavioral scenario derived from a use case, and can be instantiated as a test case:
The number of CFG paths serves as a proxy for the number of test cases. 
Based on this, we define two discrepancy-based metrics using CFG path counts.

\begin{itemize}
    \item \textbf{Discrepancy Rate:}
    The \textit{Discrepancy Rate} measures the proportion of use cases in which the number of test paths generated from the LLM-derived CFGs differed from that obtained from the ground-truth CFGs. 
    \begin{equation}
    	Discrepancy Rate (\%) = 
    	\frac{N_{diff}}{N} \times 100\%,
    \end{equation}
    where $N_{diff}$ denotes the number of use cases with mismatched path counts, and $N$ represents the total number of use cases.
    A lower discrepancy rate indicates greater consistency between the generated and reference path enumerations, suggesting behavioral alignment in the underlying CFG structures.

    \item \textbf{Average Absolute Difference (Avg.~$|\Delta|$):} 
    We computed the \textit{Average Absolute Difference ($Avg.|\Delta|$)} across all use cases.
    This quantifies the magnitude of deviation in the number of generated test paths:
    
    \begin{equation}
    	Avg.|\Delta| = \frac{1}{N} 
    	\sum_{i=1}^{N} 
    		\left| M_i^{llm} - 
    		M_i^{gt} \right|,
    \end{equation}
    where $N$ is the total number of use cases; 
    $M_i^{llm}$ is the number of test paths generated from the LLM-based CFG for the $i$-th use case; and 
    $M_i^{gt}$ is the number of paths derived from the ground-truth CFGs.
    A smaller $Avg.|\Delta|$ value indicates that the generated test paths more closely reflect the behavioral complexity captured by the ground-truth CFGs.
\end{itemize}

\textbf{LLM-based Behavioral Evaluation Metrics.}
Recent studies have shown that LLM-based evaluation can approximate human judgment in software-artifact assessment~\cite{li2024llms}.
We adopted this approach to evaluate the generated test cases against ground-truth CFG paths.
The ground-truth CFGs were manually constructed from use cases, in which each node corresponded to a statement and each path represented a complete behavioral scenario. 
Under this framework, each path served as a reference unit for evaluation.

Based on the analysis of generated test cases, we characterized the alignment between generated test cases and ground-truth paths into five representative patterns:
\begin{itemize}
    \item 
    \textbf{Exact Match ($1:1$):} 
    A path was fully covered by a single test case, representing an ideal generation.
    
    \item 
    \textbf{Fragmentation ($1:m$):} 
    A path was fully covered only by combining multiple test cases, indicating behavioral fragmentation and potential redundant execution steps.
    
    \item 
    \textbf{Uncovered Path:} 
    A ground-truth path was not fully covered by any test case, or combination of test cases, indicating missing behavior.
    
    \item 
    \textbf{Noise:} 
    A test case could not be mapped to any valid path, representing irrelevant or hallucinated behavior.
    
    \item 
    \textbf{Redundancy:} 
    Multiple test cases covered identical or highly overlapping behaviors, leading to unnecessary duplication.
\end{itemize}

\begin{figure}
    \centering
    \includegraphics[width=0.95\linewidth]{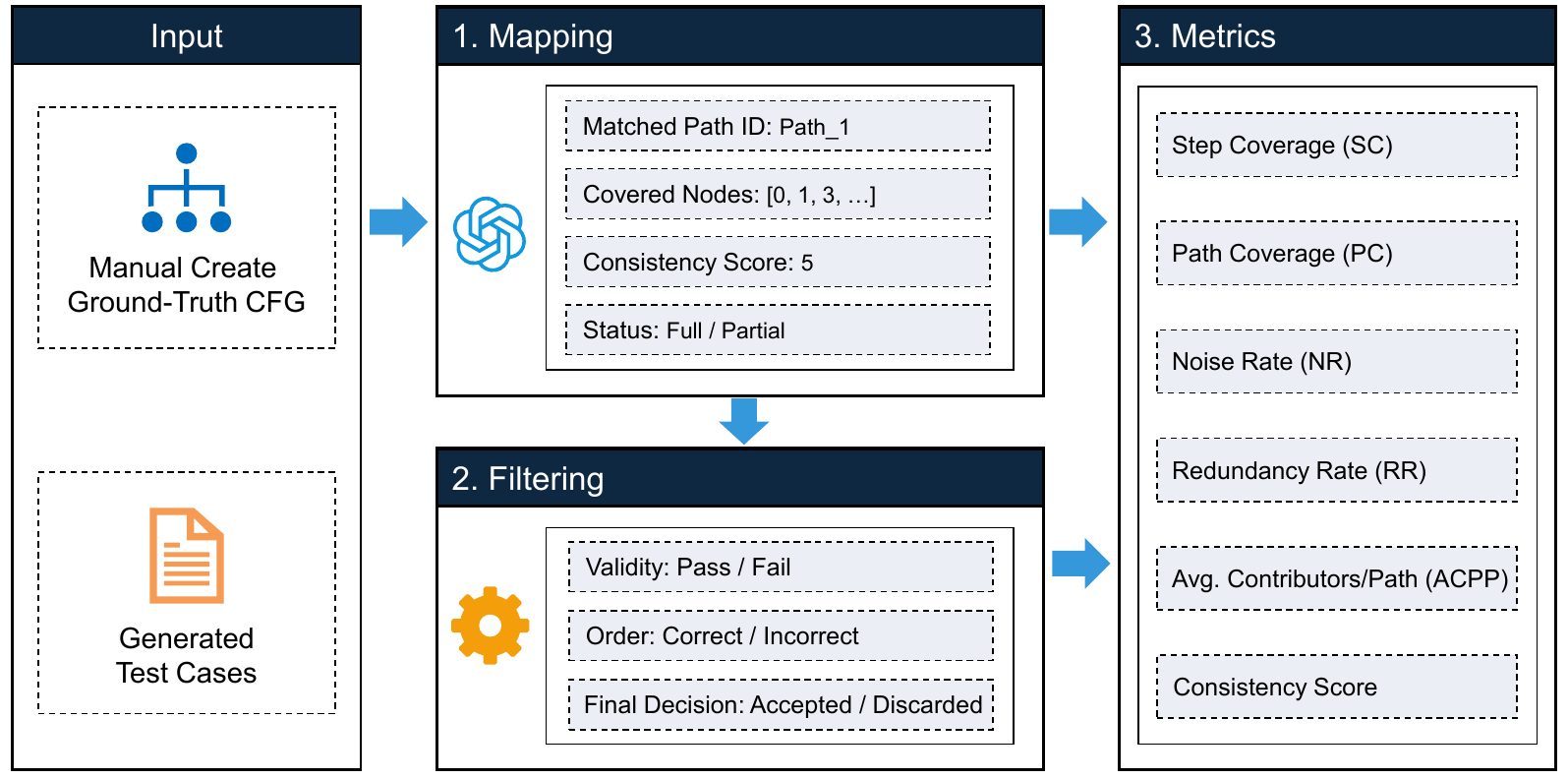}
    \caption{LLM-based behavioral evaluation pipeline.}
    \Description{Each generated test case is mapped to a ground-truth path using an LLM, followed by filtering and aggregation to compute coverage and quality metrics.}
    \label{fig:llm_eval}
\end{figure}

Figure~\ref{fig:llm_eval} shows the assessment as a three-stage automated pipeline:
\begin{enumerate}
    \item 
    \textbf{Mapping:} 
    The LLM maps each test case to the most relevant path and extracts covered nodes, completion status, and a semantic consistency score.
    
    \item 
    \textbf{Filtering:} 
    Mappings are validated using structural constraints, including the node-coverage ratio, terminal node reachability, and execution-order consistency. 
    Mappings that do not satisfy these conditions are treated as noise.    
    
    \item 
    \textbf{Aggregation:} 
    The filtered mappings are used to compute the final values of all evaluation metrics.
\end{enumerate}

Based on the alignment results from our evaluation pipeline, we defined the following metrics to quantify the behavioral consistency between the generated test cases and the ground-truth paths.
\begin{itemize}
    \item 
    \textbf{Step Coverage (SC):} 
    The proportion of unique CFG nodes in the ground truth that were covered by the generated valid test cases:
    \begin{equation}
    SC = \frac{|S_{cov}|}{|S_{gt}|},
    \end{equation}
    where $S_{gt}$ is the set of nodes in the ground-truth CFG, and $S_{cov}$ is the subset of $S_{gt}$ traversed by at least one valid test case.
    Higher values indicate better statement coverage.

    \item 
    \textbf{Path Coverage (PC):} 
    The proportion of ground-truth paths that were sufficiently covered by the generated test suite:
    \begin{equation}
        PC = \frac{N_{path\_cov}}{N_{paths}},
    \end{equation}
    where $N_{paths}$ is the total number of ground-truth paths, and $N_{path\_cov}$ is the number of paths that meet the node-coverage threshold while preserving the node order.
    Higher values indicate 
    better coverage of the ground-truth execution paths.
    
    \item 
    \textbf{Noise Rate (NR):} 
    The proportion of generated test cases that were invalid, or could not be mapped to any ground-truth path:
    \begin{equation}
        NR = \frac{N_{invalid}}{N_{TCs}},
    \end{equation}
    where $N_{invalid}$ is the number of invalid test cases, and $N_{TCs}$ represents the total number of generated test cases.
    Lower values indicate fewer irrelevant or incorrect test cases.

    \item 
    \textbf{Redundancy Rate (RR):} 
    The proportion of redundant cases among the valid test cases:
    \begin{equation}
        RR = \frac{N_{redund}}{N_{valid}},
    \end{equation}
    where $N_{redund}$ is the number of redundant test cases, and $N_{valid}$ is the number of valid test cases.
    Lower values indicate less duplication and higher efficiency.

    \item 
    \textbf{Average Contributors Per Path (ACPP):} 
    The average number of effective (non-redundant) test cases associated with each covered path:
    \begin{equation}
        ACPP = \frac{N_{valid} - N_{redund}}{N_{path\_cov}}.
    \end{equation}
    Values closer to 1.0 indicate that each path is covered by a single effective test case.

    \item 
    \textbf{Consistency:} 
    The average semantic alignment score was calculated between generated test steps and their corresponding ground-truth path statements:
    Each test step was compared with its corresponding path statement and assigned a similarity score, $t_j \in [1,5]$, reflecting the semantic correspondence.
    \begin{equation}
        Consistency = \frac{1}{N_{valid}} \sum_{j=1}^{N_{valid}} t_j.
    \end{equation}
    Higher values indicate stronger semantic alignment.
\end{itemize}

To ensure reproducibility and fair comparisons across models, all evaluations were conducted using \textit{GPT-5.4} with a fixed temperature setting of $0.0$.

\subsubsection{Human Evaluation Metrics (RQ3)}
\label{sec:satisfaction}
To assess the perceived quality of the test cases generated by \TGen\/, we conducted a practitioner study with six industry experts recruited from three different companies, each operating in a distinct domain (multimedia software, cybersecurity, and logistics). 
The experts included a senior test engineer and a quality assurance (QA) manager from each company, ensuring a balanced perspective between technical execution and quality management.
All participants had substantial professional experience in software testing, with between eight and 19 years of experience (mean: 13 years):
This ensured adequate expertise for evaluating test-case quality.

We compared the test cases generated by three approaches: 
LLM (Direct), AGORA, and \TGen\/.
All the test cases produced under identical experimental conditions. 
Because of the cost of manual evaluation, the study focused on 30 use cases (five from each dataset), for a total of 351 test cases. 
Each evaluation session lasted approximately two hours. 
Participants first received a briefing on the use cases, and then assessed the test cases independently.

Following previous work~\cite{arora2024generating, wang2025multi}, the participants rated each test case along the following four dimensions, using a five-point Likert scale (1 = Strongly Disagree; 5 = Strongly Agree):
\begin{itemize}
    \item 
    \textbf{Faithfulness:} 
    This assessed whether or not the test case preserved the intent of the original requirement or use case.  
    A relevant test case should accurately reflect the intended behavior, avoiding ambiguity, misunderstanding, or hallucinated content.

    \item 
    \textbf{Completeness:} 
    This evaluated whether or not the test case fully covered all critical behaviors, conditions, and paths described in the source use case.
    Possible situations included a complete omission of a test case or incomplete test suites (lacking essential elements such as titles, preconditions, steps, inputs, or outputs), as well as cases where full behavioral coverage was distributed across multiple test cases.

    \item 
    \textbf{Correctness:} 
    This referred to the logical and semantic validity of a test case. 
    A correct test case should be consistent with domain-specific logic, and comply with the constraints specified in the requirements. 
    Redundant steps, contradictory preconditions, or illogical action sequences 
    reduce the correctness of a test case.

    \item 
    \textbf{Clarity:} 
    This evaluated the readability and clarity (lack of ambiguity) of a test case. 
    Clear test cases present titles, preconditions, and steps in a concise and unambiguous manner, facilitating interpretation by human testers or automated tools.
\end{itemize}

We first performed a pilot study, where only a subset of the samples was evaluated by the experts:
This aimed to identify potential ambiguities in the criteria, and to assess inter-rater consistency. 
Based on the findings, we refined the guidelines and provided additional clarifications for the assessment dimensions. 
Sources of disagreement, such as fragmented test cases and overly verbose descriptions, were also addressed. 
The final evaluations were performed using the refined guidelines, providing a more consistent interpretation. 
Finally, we also collected qualitative feedback on the generated test cases.

\subsection{Implementation Details
\label{sec:implementation}}
The experiments were conducted on a Windows 10 workstation using Python 3.12.4 \cite{python3124}. 
Dependency management was handled using Conda 24.5.0 \cite{anaconda2450}, which maintained a consistent runtime environment across all tests.

The four LLMs were accessed through their official APIs: 
\textit{GPT-5.4} \cite{openai_gpt54_2026};
\textit{Claude Opus 4.6} \cite{anthropic_claude_opus46_2026}; 
\textit{Gemini 3 Flash} \cite{google_gemini3flash_2026}; and
\textit{Qwen 3.5 Plus} \cite{alibaba_qwen35plus_2026}.
To ensure deterministic outputs, all models were invoked with the temperature set to $0.0$, and all other hyperparameters kept at their default API configurations.
The \textit{Qwen 3.5 Plus} \texttt{enable\_thinking} option was disabled because it substantially increased model response time.
Preliminary experiments showed that enabling this option had negligible impact on the generated test cases, while disabling it significantly improved execution efficiency and reduced runtime disparities across the four evaluated LLMs.

To ensure evaluation consistency, we used \textit{GPT-5.4} to assess all methods in RQ2.
To support reproducibility, all datasets, configuration files, and evaluation scripts have been made publicly available on GitHub:
\url{https://github.com/aimeenorwind/\TGen/}.

\section{Experimental Results
\label{sec:experimentalresults}}

This section presents the experimental findings and provides detailed answers to the above four RQs (Section~\ref{section:RQs}).

\subsection{Answer to RQ1: How Accurate are the LLM-Generated CFGs compared to a Manually-Constructed Ground Truth?}

To evaluate the accuracy of LLM-constructed CFGs, we used node- and edge-level precision, recall, and F1-scores based on semantic matching.
We also used nGED to assess structural similarity. 
(The formal definitions of these metrics were provided in Section~\ref{section:RQ1_eval}.)

\begin{table}
    \centering
    \renewcommand{\arraystretch}{1.25}
    \setlength\tabcolsep{2.2mm}
    \caption{Accuracy of LLM-Generated CFGs Compared to the Ground-Truth CFGs,  across Datasets}
    \label{tab:rq1_cfg_accuracy}
    \footnotesize
    \begin{tabular}{c|rrr|rrr|r}
        \hline
        \multirow{2}{*}{\textbf{Model}}
        & \multicolumn{3}{c|}{\textbf{Nodes}} 
        & \multicolumn{3}{c|}{\textbf{Edges}} 
        & \multirow{2}{*}{$nGED\downarrow$} \\
        
        \cline{2-7}
        & $Precision\uparrow$ & $Recall\uparrow$ & $F1\text{-}score\uparrow$
        & $Precision\uparrow$ & $Recall\uparrow$ & $F1\text{-}score\uparrow$
        & \\
        
        \hline
        
        GPT-5.4    & \textbf{94.78} & \textbf{95.00} & \textbf{94.39} & \textbf{82.52} & 82.71 & \textbf{82.34} & \textbf{0.11} \\
        Claude Opus 4.6    & 94.25 & 94.33 & 93.90 & 82.41 & \textbf{82.99} & 82.10 & 0.12 \\
        Gemini 3 Flash & 87.15 & 88.40 & 87.28 & 72.70 & 73.05 & 72.30 & 0.20 \\
        Qwen 3.5 Plus  & 93.26 & 90.87 & 91.33 & 82.20 & 80.15 & 79.91 & 0.14 \\
        
        \hline
    \end{tabular}
\end{table}

Table~\ref{tab:rq1_cfg_accuracy} summarizes the average scores of the four LLMs across all datasets, with each row reporting the mean metric values for a specific model over all use cases. 
Overall, all models had high structural similarity, with the generated graphs closely approximating the ground truth.

\textit{GPT-5.4}, \textit{Claude Opus 4.6}, and \textit{Qwen 3.5 Plus} all had node-level F1-scores above $0.9$, and edge-level F1-scores above $0.8$. 
\textit{GPT-5.4} and \textit{Claude Opus 4.6} had comparable top-tier results, with \textit{Qwen 3.5 Plus} following closely, with a consistently strong performance. 
Notably, the edge-level metrics were consistently lower than the node-level one, across all models:
This is expected with graph construction, as an error in a single node typically impacts on multiple incident edges. 
The strong structural alignment was further supported by the nGED values remaining below $0.2$ for the leading models. 
\textit{Gemini 3 Flash} had the lowest average scores.

\begin{figure}
    \centering
    \includegraphics[width=0.9\textwidth]{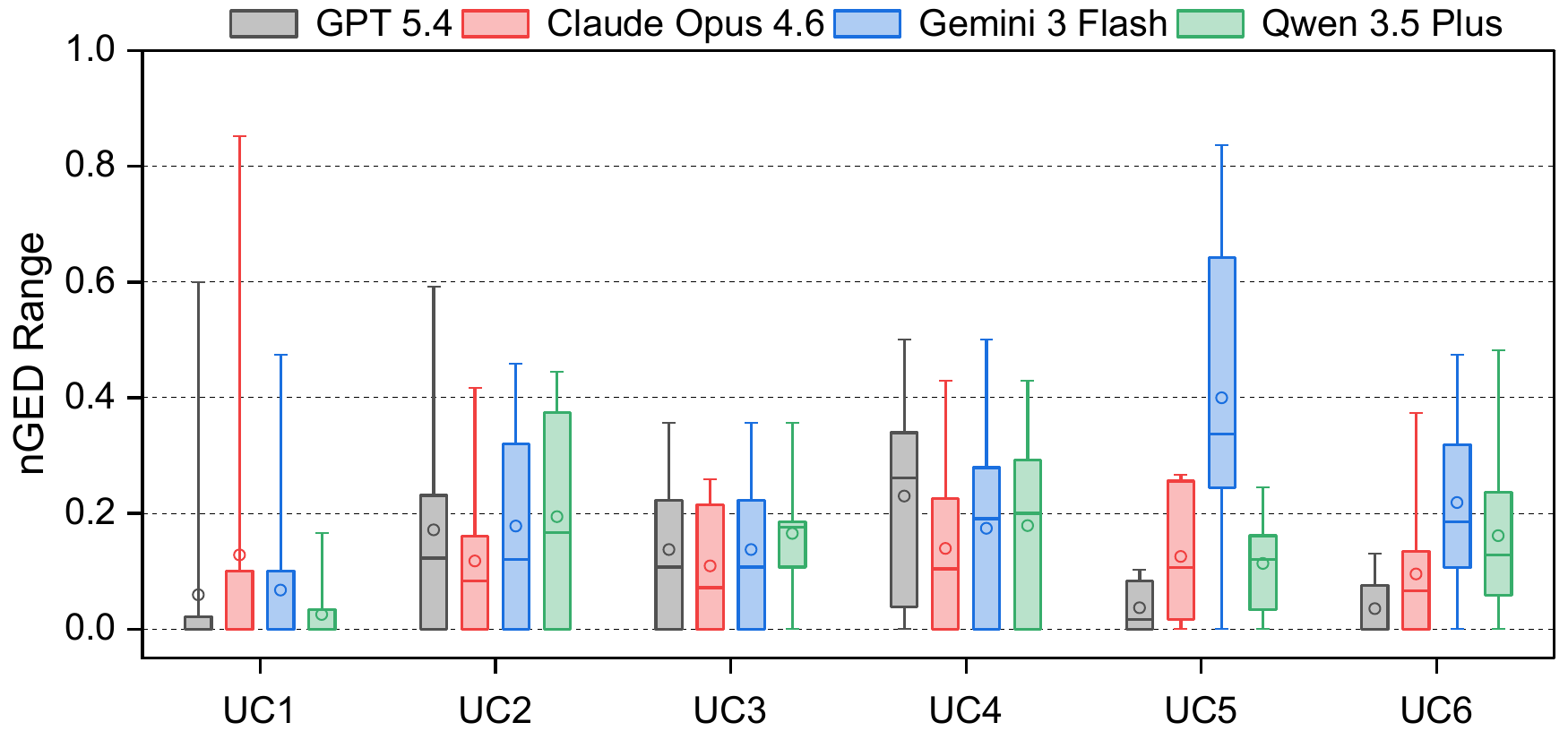}
    \caption{Distribution of nGED for CFGs generated by different LLMs.}
    \Description{Grouped box plots of nGED scores for CFGs generated by different LLMs across datasets.}  
    \label{fig:RQ1}
\end{figure}

Figure~\ref{fig:RQ1} shows the distribution of nGED scores across datasets, including the mean values, and the minimum-to-maximum ranges. 
\textit{Claude Opus 4.6} had consistently strong results across all use cases. 
While \textit{GPT-5.4} performed very well with UC1, UC5, and UC6, it had slightly higher nGED values on UC2, UC3, and UC4 compared to \textit{Claude Opus 4.6}. 
\textit{Qwen 3.5 Plus} had stable results on UC1, UC3, and UC5, with nGED values mostly below $0.2$. 
In contrast, \textit{Gemini 3 Flash} had a noticeable performance drop on UC5:
A manual inspection revealed that this was primarily caused by the omission or misplacement of a small number of edges, which altered the extracted execution paths.

\begin{tcolorbox}[breakable,colframe=black,colback=white,arc=0mm,left={1mm},top={1mm},bottom={1mm},right={1mm},boxrule={0.25mm}]
    \textit{\textbf{Summary of Answers to RQ1:}}
    The CFGs generated by \TGen\/ closely matched the ground truths, across both structural and semantic dimensions. 
    The consistently high values across all evaluation metrics
    ---
    including node-level and edge-level precision, recall, F1-score, and nGED
    ---
    show that \TGen\/ can reliably construct CFGs that are highly consistent with manually-constructed ground-truth graphs, across all the evaluated LLMs.     
\end{tcolorbox}

\subsection{Answer to RQ2: How Effectively do Test Cases Generated by \TGen\/ Align with the Ground-Truth Execution Structures, compared with the Baseline Methods?
\label{sec:RQ2}}

We evaluated the effectiveness of \TGen\/ from two perspectives: 
(1) test-case quantity alignment; and 
(2) LLM-based behavioral metrics coverage, noise, redundancy, ACPP, and consistency). 
(The detailed definitions of all the metrics were provided in Section~\ref{sec:metrics_rq2}).
In contrast to RQ3, which required practitioner evaluations, RQ2 provides a metric-driven assessment of the generated test cases compared with the ground-truth CFG execution structures.

\begin{figure}
  \centering
  \subfigure[Discrepancy rate across models and methods.]{
  \label{fig:rq2_dis}
  \Description{Grouped bar chart showing discrepancy values across models and methods.}
  \includegraphics[width=0.48\linewidth]{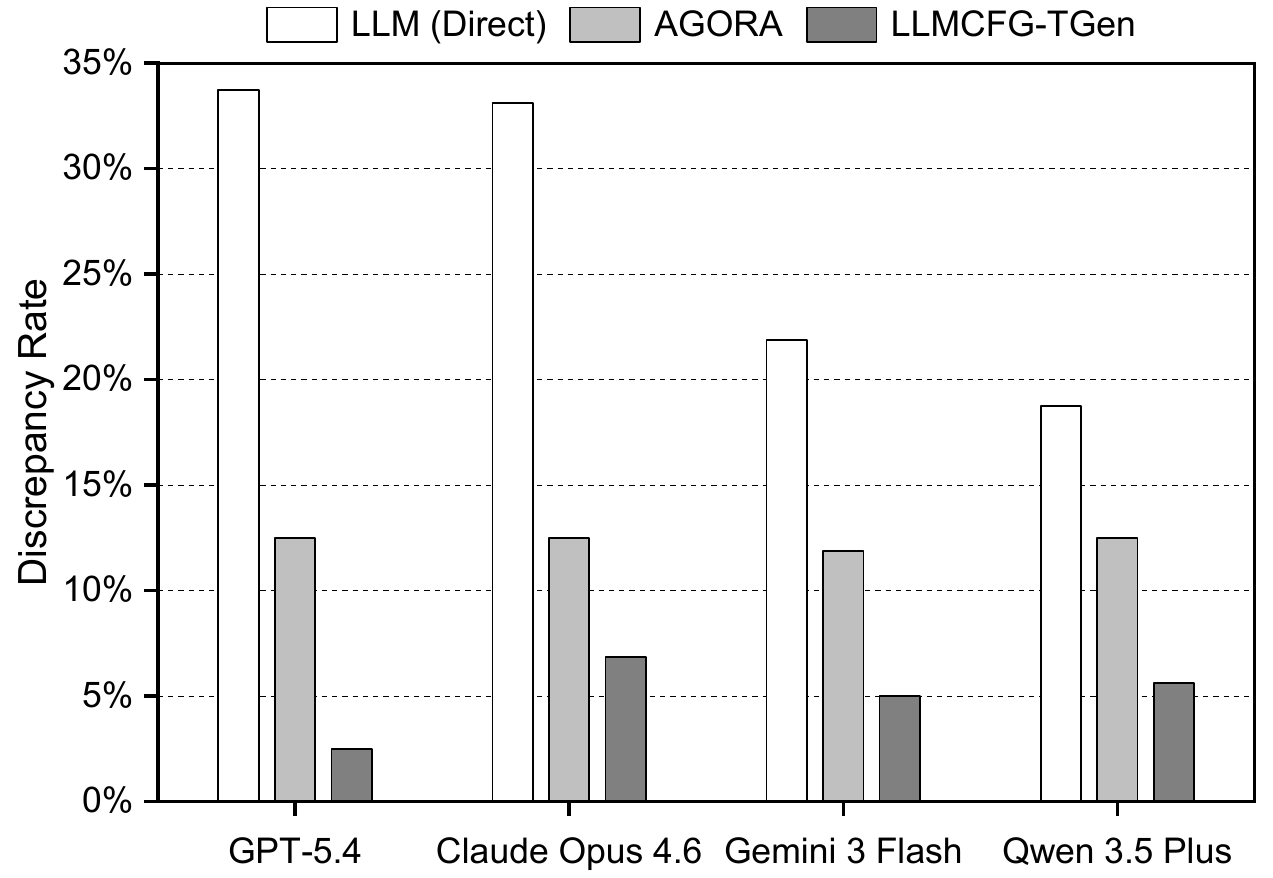}}
  \subfigure[Avg.~$|\Delta|$ across models and methods.]{
  \label{fig:rq2_avg}
  \Description{Grouped bar chart showing average absolute deviation across models and methods.}
  \includegraphics[width=0.48\linewidth]{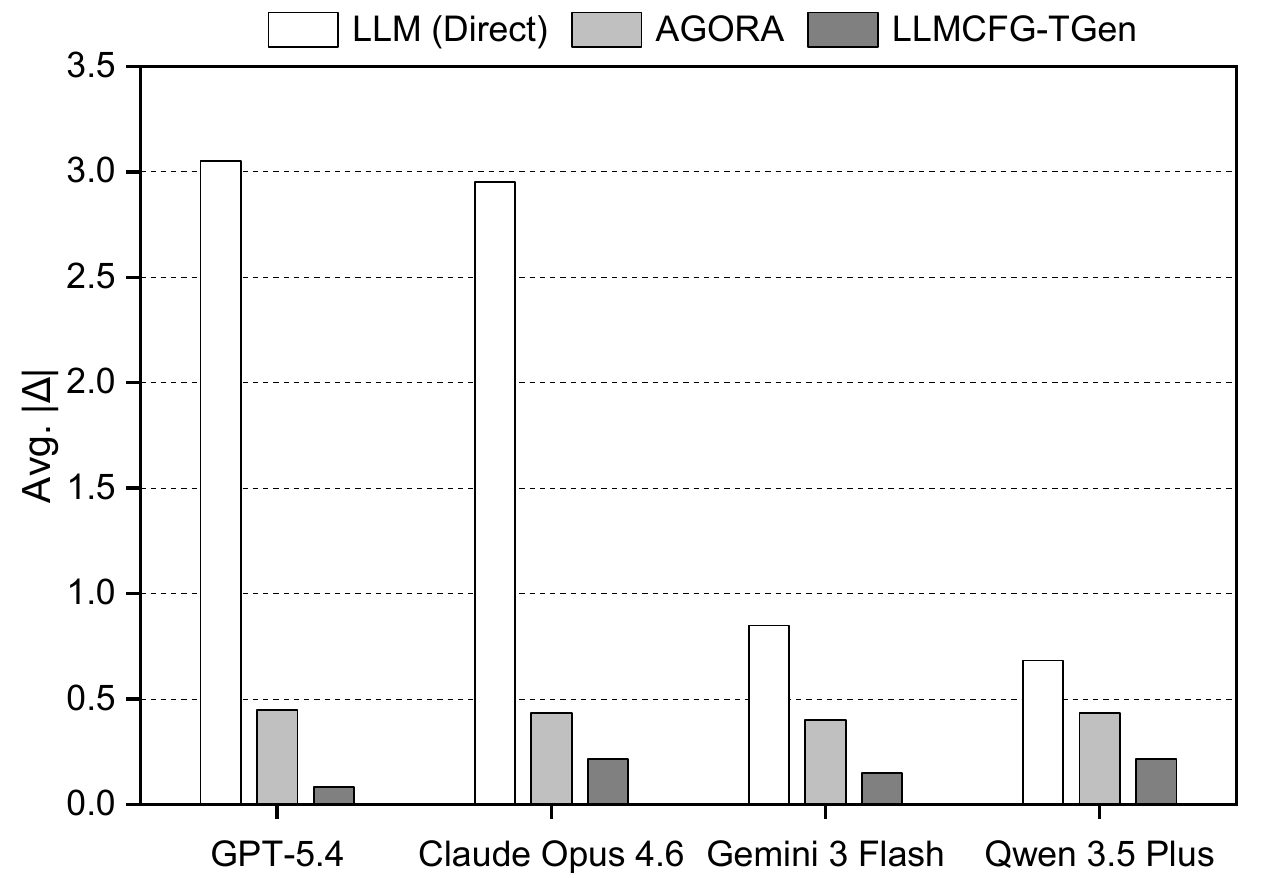}}
  \caption{Quantity deviation across LLMs.}
  \Description{Two grouped bar charts comparing three methods across four LLMs.}
  \label{fig:rq2_quantitative}
\end{figure}

\begin{table}
    \centering
    \renewcommand{\arraystretch}{1.25}
    \setlength\tabcolsep{2.4mm}
    \caption{Quantitative Comparison of Test-Case Generation Approaches across Different LLMs}
    \label{tab:rq2_quantity}
    \footnotesize
    \begin{tabular}{l l c c c c}
        \hline
        \textbf{Model} & \textbf{Method} & \textbf{\#Tests (\textit{vs.} 160)} & \textbf{\#Steps / \#Tests} & $Discrepancy~(\%) \downarrow$ & $Avg.|\Delta|\downarrow$ \\
        \hline
        
        \multirow{3}{*}{GPT-5.4}
        & LLM (Direct) & 339 & 2.82 & 33.75 & 3.05 \\
        & AGORA        & 155 & 4.57 & 12.50 & 0.45 \\
        & \textbf{LLMCFG-TGen} & 159 & 3.62 & \textbf{2.50} & \textbf{0.08} \\
        \hline
        
        \multirow{3}{*}{Claude Opus 4.6}
        & LLM (Direct) & 333 & 2.91 & 33.13 & 2.95 \\
        & AGORA        & 150 & 4.01 & 12.50 & 0.43 \\
        & \textbf{LLMCFG-TGen} & 161 & 4.06 & \textbf{6.88} & \textbf{0.22} \\
        \hline
        
        \multirow{3}{*}{Gemini 3 Flash}
        & LLM (Direct) & 199 & 1.87 & 21.88 & 0.85 \\
        & AGORA        & 152 & 2.33 & 11.88 & 0.40 \\
        & \textbf{LLMCFG-TGen} & 163 & 2.75 & \textbf{5.00} & \textbf{0.15} \\
        \hline
        
        \multirow{3}{*}{Qwen 3.5 Plus}
        & LLM (Direct) & 183 & 3.01 & 18.75 & 0.68 \\
        & AGORA        & 156 & 3.49 & 12.50 & 0.43 \\
        & \textbf{LLMCFG-TGen} & 161 & 3.63 & \textbf{5.63} & \textbf{0.22} \\
        \hline
        
    \end{tabular}
\end{table}

\textbf{Quantity Alignment.}
Figure~\ref{fig:rq2_quantitative} presents the discrepancy and Avg.~$|\Delta|$ across different models and methods. 
As shown in Figures~\ref{fig:rq2_dis} and~\ref{fig:rq2_avg}, \TGen\/ consistently achieved the lowest discrepancy and deviation across all evaluated LLMs:
This indicates that the number of generated test cases closely matched the ground-truth CFG paths. 
In contrast, LLM (Direct) tended to significantly over-generate test cases, and AGORA had moderate deviation from the ground truth.

Table~\ref{tab:rq2_quantity} reports the total number of generated test cases, average steps per test case, and the corresponding discrepancy metrics. 
The results confirm that \TGen\/ produced test suites with sizes the closest to the ground truth (160 paths):
This shows superior alignment with the underlying CFG structure.  
LLM (Direct) tended to generate test cases with fewer steps (many consisting of only a single step): 
This indicates fragmented and less informative test scenarios. 
This behavior was not observed in AGORA or \TGen\/, both of which produced more coherent multi-step test cases.

\begin{table}
    \centering
    \renewcommand{\arraystretch}{1.25}
    \setlength\tabcolsep{2.9mm}
    \caption{LLM-Based Assessment of Generated Test Cases}
    \label{tab:rq2_llm_assessment}
    \footnotesize
    \begin{tabular}{l|l|rr|rr|c|c}
        \hline
        \multirow{2}*{\textbf{Model}} & \multirow{2}*{\textbf{Method}} 
        & \multicolumn{2}{c|}{\textbf{Coverage (\%)}} 
        & \multicolumn{2}{c|}{\textbf{Correctness (\%)}} 
        & \multirow{2}*{$ACPP\downarrow$} 
        & \multirow{2}*{$Consistency \uparrow$} \\
        
        \cline{3-4} \cline{5-6}
        & 
        & $SC\uparrow$ & $PC\uparrow$
        & $NR\downarrow$ & $RR\downarrow$ 
        & 
        &  \\
        
        \hline
        
        \multirow{3}{*}{GPT-5.4}
        & LLM (Direct) & 89.18 & 78.30 & 2.97 & 29.28 & 1.88 & 4.48 \\
        & AGORA        & 84.35 & 76.87 & \textbf{0.00} & 1.41 & 1.34 & 4.59 \\
        & \textbf{LLMCFG-TGen} & \textbf{93.60} & \textbf{91.09} & \textbf{0.00} & \textbf{0.00} & \textbf{1.09} & \textbf{4.81} \\
        \cline{1-8}
        
        \multirow{3}{*}{Claude Opus 4.6}
        & LLM (Direct) & \textbf{94.05} & \textbf{89.64} & 2.12 & 36.43 & 1.52 & 4.74 \\
        & AGORA        & 80.88 & 74.30 & 4.70 & \textbf{2.65} & 1.23 & 4.85 \\
        & \textbf{LLMCFG-TGen} & 93.37 & 89.38 & \textbf{0.95} & 3.10 & \textbf{1.08} & \textbf{4.91} \\
        \cline{1-8}
        
        \multirow{3}{*}{Gemini 3 Flash}
        & LLM (Direct) & 84.59 & 73.91 & 0.35 & 13.80 & 1.86 & 4.75 \\
        & AGORA        & 81.81 & 73.85 & 1.11 & 4.87 & 1.19 & 4.84 \\
        & \textbf{LLMCFG-TGen} & \textbf{93.18} & \textbf{91.26} & \textbf{0.00} & \textbf{2.55} & \textbf{1.09} & \textbf{4.92} \\
        \cline{1-8}
        
        \multirow{3}{*}{Qwen 3.5 Plus}
        & LLM (Direct) & 87.34 & 80.78 & \textbf{0.35} & 11.47 & 1.38 & 4.76 \\
        & AGORA        & 84.48 & 78.93 & 1.39 & \textbf{4.52} & 1.26 & 4.75 \\
        & \textbf{LLMCFG-TGen} & \textbf{90.37} & \textbf{86.83} & 0.69 & 7.67 & \textbf{1.11} & \textbf{4.85} \\
        \hline
        
        \multicolumn{8}{l}{\textit{SC}: Step Coverage; \textit{PC}: Path Coverage; \textit{NR}: Noise Rate; \textit{RR}: Redundancy Rate; and \textit{ACPP}: Average Contributors Per Path}
    \end{tabular}
\end{table}

\textbf{LLM-based Behavioral Quality.}
We further evaluated the generated test cases using an LLM-based assessment. 
Table~\ref{tab:rq2_llm_assessment} reports the average performance across all datasets in terms of coverage, correctness, efficiency, and semantic consistency.  

The results show that \TGen\/ had the highest Step and Path Coverage (SC and PC), in almost all evaluated LLMs
---
the only exception was \textit{Claude Opus 4.6}, for which LLM (Direct) had slightly higher coverage (94.05\%, compared with \TGen\/'s 93.37\%), but with an associated increased noise and redundancy.

Trends for the correctness-related metrics were less uniform across the models: 
\TGen\/ achieved the best noise and redundancy rates (NR and RR) on \textit{GPT-5.4} and \textit{Gemini 3 Flash};
AGORA had the lowest redundancy with \textit{Claude Opus 4.6} (2.65\%, compared with 3.10\% for \TGen\/);
LLM (Direct) had the lowest noise with \textit{Qwen 3.5 Plus} (0.35\%, compared with 0.69\% for \TGen\/); and 
AGORA achieved the lowest redundancy with \textit{Qwen 3.5 Plus} (4.52\%, compared with 7.67\% for \TGen\/). 
Overall, \TGen\/ had strong performance across correctness-related dimensions, outperforming the baselines in most settings. 
Notably, LLM (Direct) generally had higher redundancy rates across models, indicating less compact test generation.
In terms of efficiency, \TGen\/ consistently had the lowest ACPP across all models, indicating the most compact test generation per path.
Finally, all three methods achieved relatively high semantic consistency scores, indicating that the LLM-generated test cases generally preserved the intent of the original requirements. 
Overall, \TGen\/ had the highest consistency across models, demonstrating the strongest alignment with ground-truth test semantics.

\begin{figure}
  \centering
  \includegraphics[width=0.9\linewidth]{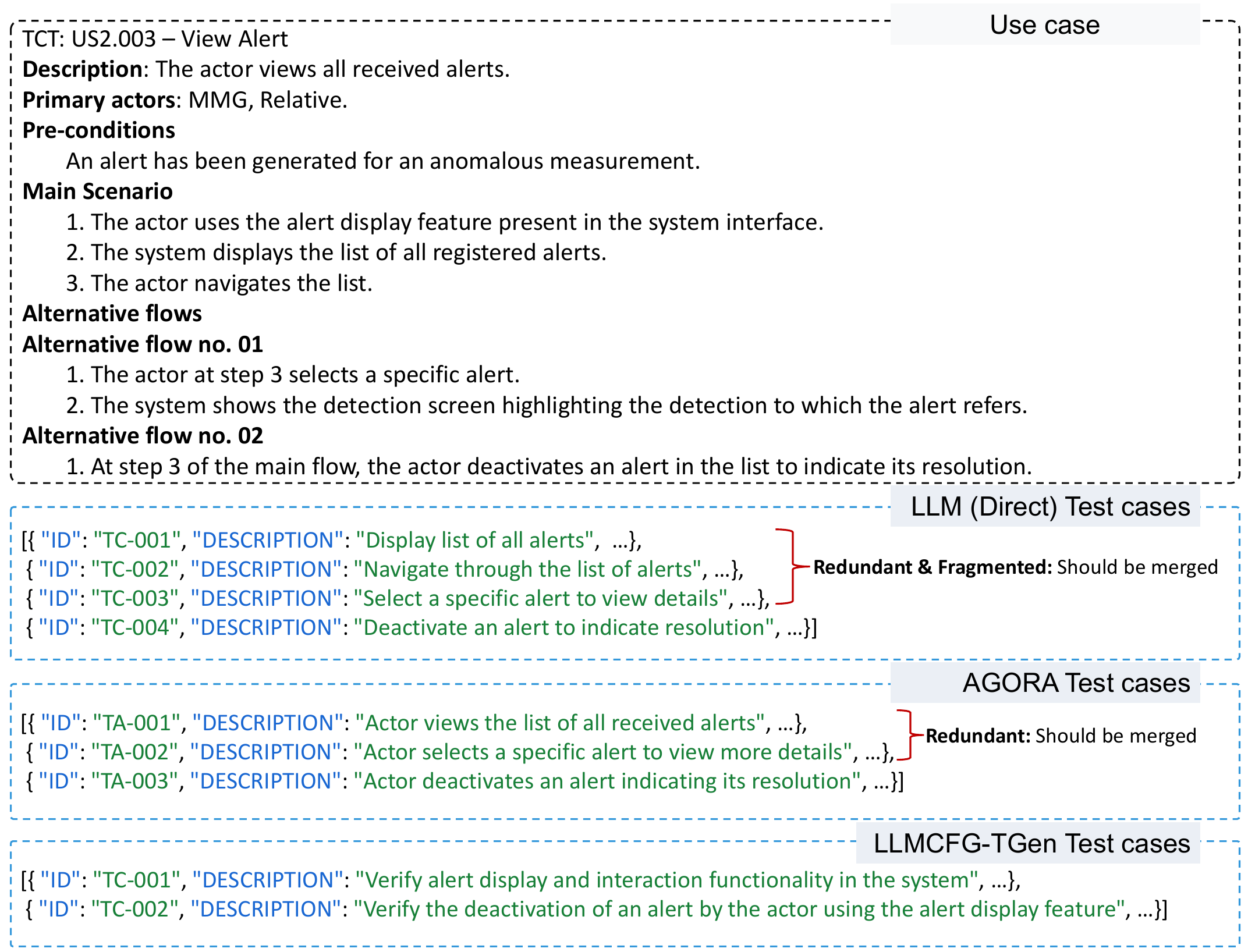}
  \caption{“View alert” use case with generated test cases.}  
  \Description{Comparison of test cases generated by three approaches.}
  \label{fig:RQ2}
\end{figure}

To better understand these differences, we analyzed those use cases with noticeable discrepancies, most of which involved complex conditional branching. 
Figure~\ref{fig:RQ2}, for example, shows the ``View alert'' use case and the corresponding test cases generated by \textit{GPT-5.4}.
In this example, LLM (Direct) generated four separate test cases to achieve full coverage. 
However, the combination of TC-001, TC-002, and TC-003 together describe a single coherent execution flow, leading to both redundancy and fragmentation in the generated test suite.
AGORA, which relies on the structured format of use cases, did not capture the relationships between the main scenarios and the alternative flows, treating them as independent paths, and producing redundant test cases.
In contrast, \TGen\/, which uses CFG representations to interpret conditions expressed in NL, was able to precisely identify valid execution paths, and thus generate complete test cases without redundancy. 
These conditional branching structures often appear in forms such as ``If \ldots then \ldots otherwise \ldots'' or ``In a solitaire game, \ldots; In a two-player game, \ldots''.
\TGen\/ consistently identified these branch structures correctly, resulting in more precise and coherent test cases.

\begin{tcolorbox}[breakable,colframe=black,colback=white,arc=0mm,left={1mm},top={1mm},bottom={1mm},right={1mm},boxrule={0.25mm}]
    \textit{\textbf{Summary of Answers to RQ2:}}
    \TGen\/ consistently outperformed the baseline methods across all evaluated dimensions, achieving a better balance among test-case quantity alignment, behavioral coverage, efficiency, and semantic consistency. 
    In terms of quantity alignment, \TGen\/ produced test suites whose sizes were closest to the ground-truth CFG paths, achieving the lowest discrepancy and average absolute deviation.
    Regarding the behavioral quality, \TGen\/ maintained high coverage while effectively reducing noise and redundancy, resulting in more coherent test suites. 
    It also produced more compact test cases (lowest ACPP), and achieved the highest semantic consistency across all models, indicating reduced fragmentation and strong alignment with ground-truth semantics.
    In contrast, LLM (Direct) tended to generate a large number of test cases, leading to redundancy and fragmented scenarios;
    AGORA partially mitigated this effect, but still had lower coverage, as well as residual noise and redundancy.
\end{tcolorbox}

\subsection{Answer to RQ3: How Does the Quality of \TGen\/ Test Cases compare with the Baseline Methods?
\label{sec:RQ3}}

To evaluate the practical utility of the generated test cases, we conducted a study involving six industry practitioners recruited from three companies across different industry domains.
In contrast to RQ2, which provided a metric-driven evaluation against ground-truth CFG execution structures, RQ3 assessed the {\em perceived} quality of the generated test cases, from a practitioner's perspective, focusing on usability and 
semantic quality.
The participants assessed the test cases generated by LLM (Direct), AGORA, and \TGen\/ using 30 use cases 
--- 
the first five use cases from each dataset
---
for a total of 351 evaluated test cases.
The evaluation was based on four criteria: \textit{faithfulness};
\textit{completeness};
\textit{correctness}; and 
\textit{clarity}.
(Their detailed descriptions were provided in Section~\ref{sec:satisfaction}). 

\begin{figure}
    \centering
    \includegraphics[width=0.65\linewidth]{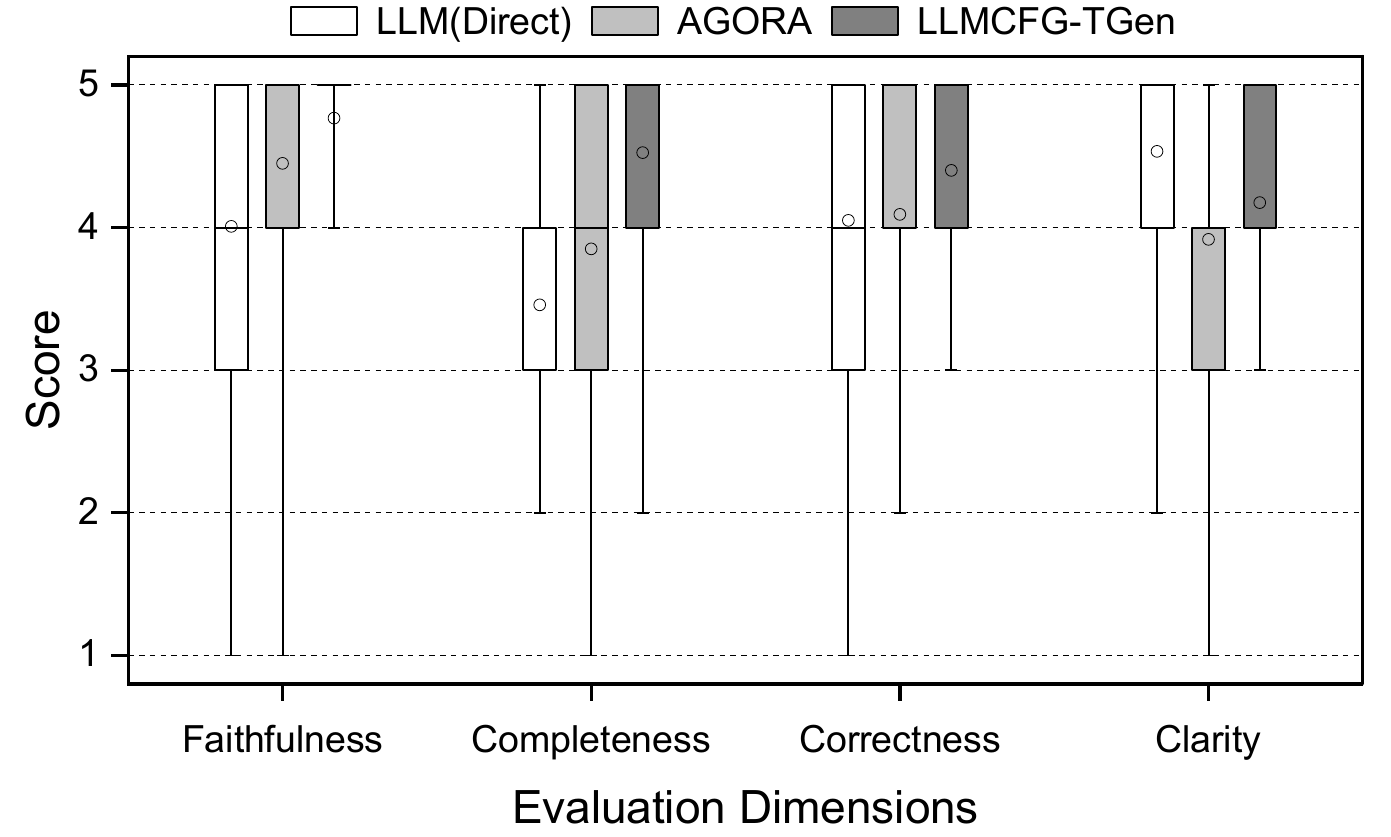}
    \caption{Practitioner Evaluation of Generated Test Cases.}
    \Description{Group plot of nGED distributions across CFGs generated by four LLMs, indicating structural similarity to ground truth.}  
    \label{fig:RQ3_plot}
\end{figure}

Figure~\ref{fig:RQ3_plot} shows the distribution of practitioner scores across all evaluation dimensions.
TGen\/ had the highest median \textit{faithfulness} scores
---
LLM (Direct) and AGORA more frequently introduced hallucinated or unsupported steps, with the issue being more evident in LLM (Direct).
\TGen\/ received consistently higher \textit{completeness} ratings, indicating better coverage of business scenarios than the baselines.
All three approaches had similarly \textit{correctness} scores, with \TGen\/ showing a slight advantage. 
The practitioner feedback suggested that correctness was often influenced by step readability, with less explicit descriptions potentially being negatively perceived. 
LLM (Direct) received higher \textit{clarity} ratings due to its concise and direct generation style:
\TGen\/ focuses on full execution paths, which may improve structural completeness but can reduce perceived conciseness in more complex cases; and
AGORA received the lowest \textit{clarity} ratings, mainly due to verbose step descriptions and overly detailed preconditions.

We conducted a statistical analysis to evaluate the significance of the observed differences. 
The Wilcoxon signed-rank test~\cite{wilcoxon1945individual} was used to compute $p$-values for pairwise comparisons between methods:
This was chosen due to its suitability for paired, non-normally distributed data \cite{huang2024toward}. 
To quantify the magnitude of differences between methods, we also computed the Vargha-Delaney effect size ($\hat{A}_{12}$)~\cite{vargha2000critique}:
This has been widely used in empirical software engineering studies \cite{huang2024toward}.  
For two compared methods, $\mathcal{M}_1$ and $\mathcal{M}_2$, the effect size is defined as:
\begin{equation}
    \hat{A}_{12}(\mathcal{M}_1, \mathcal{M}_2)
    =
    \frac{R_1 / P - (P + 1)/2}{Q},
\end{equation}
where $R_1$ is the rank sum of $\mathcal{M}_1$ (i.e., the sum of ranks assigned to its observations after pooling all samples from both methods); 
and $P$ and $Q$ denote the numbers of observations in $\mathcal{M}_1$ and $\mathcal{M}_2$, respectively. 
An $\hat{A}_{12}$ value of $0.50$ indicates no difference between the two methods;
$\hat{A}_{12}(\mathcal{M}_1, \mathcal{M}_2) > 0.50$ indicates that $\mathcal{M}_1$ performs better than $\mathcal{M}_2$; and
$\hat{A}_{12}(\mathcal{M}_1, \mathcal{M}_2) < 0.50$ indicates that $\mathcal{M}_2$ performs better than $\mathcal{M}_1$. 
Kendall’s coefficient of concordance ($W$)~\cite{kendall1939problem} was also used to measure the level of agreement among practitioners, across different use cases, for each evaluation dimension.

\begin{table}
    \centering
    \renewcommand{\arraystretch}{1.25}
    \setlength\tabcolsep{3.5mm}
    \caption{Statistical Analysis Results for each Evaluation Dimension}
    \label{tab:rq3_stats}
    \footnotesize
    \begin{tabular}{l|cc|cc|c}
        \hline
        
        \textbf{Dimension}
        & \multicolumn{2}{c|}{\textbf{\TGen\/ \textit{vs.} LLM (Direct)}}
        & \multicolumn{2}{c|}{\textbf{\TGen\/ \textit{vs.} AGORA}}
        & \multirow{2}*{\textbf{Kendall's $W$}} \\
        
        \cline{2-5}
        
        & $p$-value & Effect size ($\hat{A}_{12}$)
        & $p$-value & Effect size ($\hat{A}_{12}$)
        & \\
        
        \hline
        
        Faithfulness & $4.08 \times 10^{-11}$ & 0.7254 & $7.89 \times 10^{-6}$  & 0.5918 & 0.1197 \\
        Completeness & $1.29 \times 10^{-13}$ & 0.8183 & $5.14 \times 10^{-8}$  & 0.6985 & 0.2327 \\
        Correctness  & $5.21 \times 10^{-4}$  & 0.5921 & $3.67 \times 10^{-4}$  & 0.5968 & 0.0369 \\
        Clarity      & $2.46 \times 10^{-4}$  & 0.3644 & $1.03 \times 10^{-2}$  & 0.5852 & 0.1147 \\
        
        \hline
    \end{tabular}
\end{table}

Table~\ref{tab:rq3_stats} summarizes the results.
All the $p$-values, across all the evaluation dimensions, were below the $0.05$ threshold:
This confirms that the performance differences between \TGen\/ and the baselines were statistically significant.
Regarding the effect size, \TGen\/ demonstrated a clear advantage in \textit{faithfulness}, \textit{completeness}, and \textit{correctness}. 
This included an $\hat{A}_{12}$ of $0.8183$ compared with LLM (Direct), indicating a strong practical effect. 
The trend for \textit{clarity}, however, was different:
\TGen\/ had an $\hat{A}_{12}$ of $0.3644$ ($< 0.5$) against LLM (Direct). 
This suggests that unconstrained generation may be more intuitive than \TGen\/'s structured synthesis. 
Finally, the Kendall’s $W$ values ranged from $0.0369$ to $0.2327$, indicating low to moderate agreement among practitioners across dimensions, which is common in subjective human evaluations~\cite{kendall1939problem}.

In addition to the quantitative evaluation, we also collected qualitative feedback from practitioners regarding their overall impressions of the generated test cases.
LLM (Direct) was generally considered easier to read, due to its concise generation style. 
However, practitioners also noted that its outputs were often overly fine-grained, which may reduce their practicality for real-world test-design tasks.
AGORA was reported to produce more structured and business-oriented test cases. 
However, practitioners pointed out that its outputs may become less stable when handling complex use cases involving 
multiple interacting branches and conditional decision points.
\TGen\/ was considered to have the strongest alignment with the original requirements, and the most complete end-to-end scenario coverage. 
However, some generated test cases were considered overly concise, which occasionally reduced readability in more complex scenarios.
The practitioners also emphasized that the increasing adoption of LLMs in requirements engineering and software development was making software testing more challenging. 
They highlighted the need for automated testing support that would not only generate effective test cases, but could also transform them into executable test scripts integrated with the target system.

\begin{tcolorbox}[breakable,colframe=black,colback=white,arc=0mm,left={1mm},top={1mm},bottom={1mm},right={1mm},boxrule={0.25mm}]
\textit{\textbf{Summary of Answers to RQ3:}}
    The evaluation results showed that \TGen\/ consistently outperformed both baselines in terms of \textit{faithfulness}, \textit{completeness}, and \textit{correctness}, with statistically significant differences. 
    The \TGen\/ \textit{clarity} scores were higher scores than AGORA's, but  lower than LLM (Direct):
    This indicated that structural constraints may impact the perceived readability of the generated outputs.
    Overall, the practitioners regarded \TGen\/ as a more reliable and practical tool for business-oriented testing.
    They also highlighted that future work may include transforming the abstract test cases into executable scripts.
\end{tcolorbox}

\subsection{Answer to RQ4: How Do Different LLMs Compare in Terms of Cost-Effectiveness?
\label{sec:RQ4}}

\begin{figure}
  \centering
  \subfigure[Token consumption.]{
  \label{fig:cost_token}
  \Description{Clustered bar chart comparing total token usage for different models under different methods.}
  \includegraphics[width=0.48\linewidth]{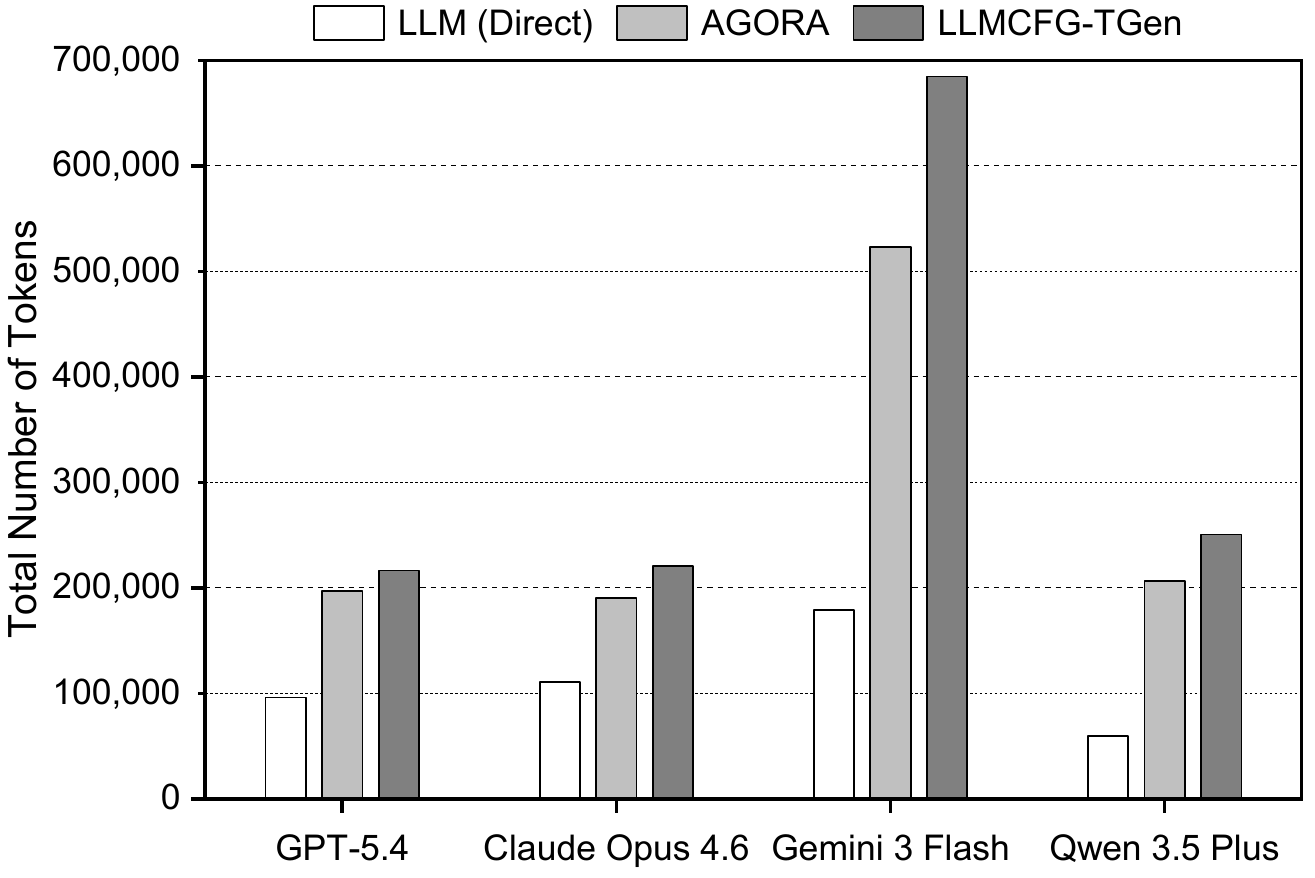}}
  \subfigure[Execution time.]{
  \label{fig:cost_time}
  \Description{Clustered bar chart comparing total execution time for different models under different methods.}
  \includegraphics[width=0.48\linewidth]{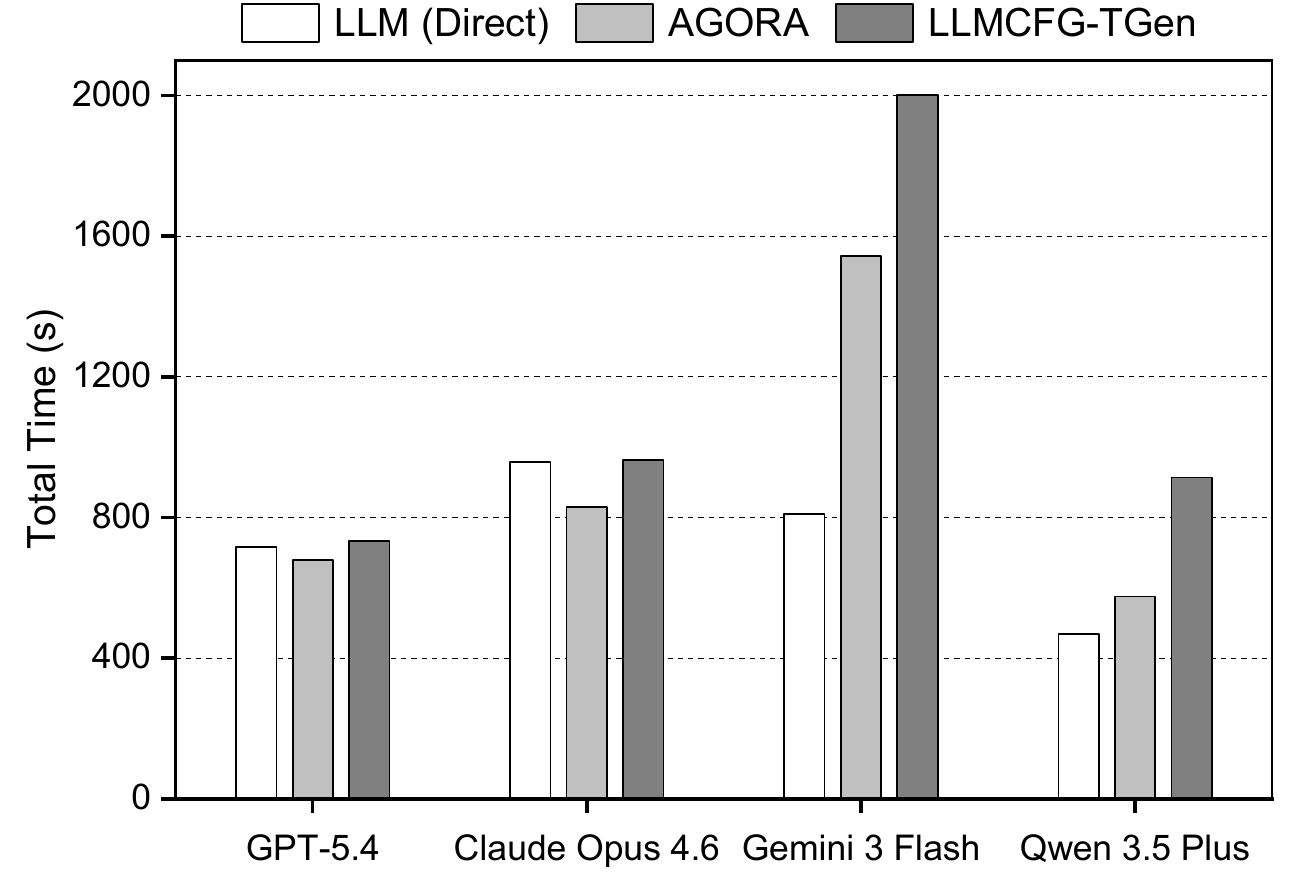}}
  \caption{Cost comparison across models and methods.}
  \Description{Comparison of computational cost across different LLMs and methods.}
  \label{fig:trend}
\end{figure}

To assess the impact of different LLMs on generation efficiency, we conducted a comparative study across the four representative models (\textit{GPT-5.4} \cite{openai_gpt54_2026}, \textit{Claude Opus 4.6}~\cite{anthropic_claude_opus46_2026}, \textit{Gemini 3 Flash}~\cite{google_gemini3flash_2026}, and \textit{Qwen 3.5 Plus}~\cite{alibaba_qwen35plus_2026}).
Figures~\ref{fig:cost_token} and~\ref{fig:cost_time} present the total token consumption and execution times across the different models and methods.

Figure~\ref{fig:cost_token} shows that LLM (Direct) consistently required the fewest tokens, and \TGen\/ required the most:
This is due to \TGen\/'s additional reasoning and structural-validation steps. 
\textit{GPT-5.4} and \textit{Claude Opus 4.6} had similar scaling patterns across methods, but \textit{Gemini 3 Flash} had a pronounced increase in token usage under AGORA and \TGen\/:
This indicates a higher sensitivity to multi-step generation pipelines.
\textit{Qwen 3.5 Plus} had the lowest token consumption with the direct generation setting, substantially lower than the other models:
This may be partially attributable to the configuration of the \texttt{enable\_thinking} option, which we had disabled in our experiments.

Figure~\ref{fig:cost_time} shows that \textit{GPT-5.4} and \textit{Claude Opus 4.6} had relatively stable execution behavior. 
\textit{Gemini 3 Flash}, in contrast, took longer with AGORA and \TGen\/, with execution times exceeding 2000 seconds in the most complex setting:
This indicates that multi-step reasoning pipelines can reduce runtime efficiency.
Compared to the other models, \textit{Qwen 3.5 Plus} took less time under LLM (Direct) and AGORA, and its execution time remained comparable to \textit{GPT-5.4} and \textit{Claude Opus 4.6} under \TGen\/ (with all three taking less than 1000 seconds).
This efficiency may be attributed to its inference configuration, where the \texttt{enable\_thinking} option was disabled.

These observations are consistent with the data in Table~\ref{tab:rq2_quantity}, where variations in the number of generated test cases and the average number of steps per test case directly contributed to differences in token consumption and execution time.
The number of CFG regenerations remained consistently low across all models. 
Among the 60 evaluated use cases, only 2 use cases triggered CFG regeneration when using \textit{GPT-5.4}, \textit{Claude Opus 4.6}, and \textit{Gemini 3 Flash}, while \textit{Qwen 3.5 Plus} required regeneration for 4 use cases. 
This indicates that the verification-and-regeneration mechanism was triggered only infrequently.
Consequently, the effect of these loops on the overall computational overhead is marginal, and does not materially increase the total cost.

\begin{tcolorbox}[breakable,colframe=black,colback=white,arc=0mm,left={1mm},top={1mm},bottom={1mm},right={1mm},boxrule={0.25mm}]
\textit{\textbf{Summary of Answers to RQ4:}}
    \TGen\/ introduces additional computational overhead in terms of both tokens and execution time, due to its multi-step generation and validation process. 
    However, as shown in the investigations for RQ1--RQ3, this cost was accompanied by substantial improvements in structural accuracy and human-evaluated effectiveness:
    This indicates that the overhead represents a justified trade-off for better quality.
    Among the evaluated models, \textit{Qwen 3.5 Plus} was the most cost-effective, while \textit{GPT-5.4} and \textit{Claude Opus 4.6} provided a robust balance between cost and reasoning depth. 
    These improvements were consistent across all evaluated LLMs, suggesting that \TGen\/ generalizes well across different models. 
\end{tcolorbox}

\subsection{Threats to Validity}

This section discusses some possible threats to the validity of our study and findings. 
It also describes the measures taken to address them.

\subsubsection{Internal Validity} 
The results may be influenced by variability in the LLM-based evaluation and the subjectivity of human assessment. 
For the automated evaluation, to improve reproducibility, we used \textit{GPT-5.4} with fixed prompts, and a temperature setting of $0.0$. 
To reduce bias in the human study, the identities of the compared methods were anonymized. 
Furthermore, a pilot study was also conducted to align evaluators’ understanding of the assessment criteria, and to refine the evaluation guidelines.

\subsubsection{Construct Validity}
Three complementary sets of metrics were used in the study:
(1) CFG-level metrics (node/edge-level F1-score and nGED); 
(2) test-case-level quantitative and behavioral metrics (Discrepancy Rate, Avg.~$|\Delta|$, Coverage, Noise, Redundancy, ACPP, and Consistency); and 
(3) practitioner-oriented metrics (Faithfulness, Completeness, Correctness, and Clarity).
Each set has limitations:
For example, although the \textit{Discrepancy Rate} shows the number of differing use cases, it does not capture the magnitude of those differences. 
To address this limitation, we reported the results across multiple complementary metrics, providing a broader evaluation perspective, and reducing the risk of misinterpretation from any single metric.

\subsubsection{External Validity}
We also acknowledge the limitations of our dataset selection. 
Although the evaluation spans multiple domains and formats, the coverage was not exhaustive. 
Certain domain-specific structures or linguistic patterns may be underrepresented, introducing potential selection bias and limiting generalizability. 
Future work should adopt a broader sampling strategy to include a wider range of styles and domain characteristics.

\section{Discussion
\label{sec:discussion}}

This section discusses the strengths and limitations of \TGen\/ for generating test cases from NL use-case descriptions, across six multi-domain datasets, using multiple evaluation metrics.

\subsection{Strengths}
\subsubsection{Input Flexibility and Domain Independence}
Unlike traditional methods \cite{wang2020automatic, alrawashed2019automated} that depend on rigid templates or domain-specific models, \TGen\/ demonstrated high flexibility in handling diverse use-case formats.  
Across the six datasets, \TGen\/ consistently produced high-quality test cases, with minimal manual intervention, demonstrating strong robustness and adaptability to heterogeneous requirement descriptions.

\subsubsection{Automated and Verifiable Intermediate Representations}
Unlike approaches that rely on manually-constructed behavioral models, \TGen\/ automatically generates CFGs in JSON format directly from NL use-case descriptions. 
Before test-case generation, the generated CFGs are subjected to automated structural validation, ensuring completeness, reachability, and logical consistency. 
This verifiable intermediate representation improves the reliability and traceability of the overall pipeline, and substantially reduces the manual modeling effort. 
A web-based interface further improves the usability and accessibility of the approach for practical test-generation workflows.

\subsubsection{End-to-End Path Coverage and Industrial Alignment}
The experimental results showed that \TGen\/ can achieve comprehensive coverage of complex business flows, while maintaining minimal redundancy and noise. 
\TGen\/ preserved the structural integrity of the business logic within each test case, without producing the fragmented or disconnected steps often seen with LLM-Direct. 
Practitioner evaluations confirmed that this end-to-end coherence made the test suites highly representative of real-world industrial testing practices. 
The experts also noted that, while direct generation could produce more fluent NL expressions, the \TGen\/ test cases were more suitable for systematic verification, and for ensuring traceability to the structured execution paths specified in the requirements.

\subsection{Weaknesses}

\subsubsection{Scalability and Use-Case Complexity}
The current framework processes one use case at a time, and does not yet support multiple related use cases:
Future work will extend the approach to model relationships across connected use cases. 
The efficiency and scalability of path enumeration should also be enhanced to support large-scale and behaviorally-complex use cases. 
In practice, excessively large use cases containing multiple tightly coupled functionalities can significantly increase CFG complexity and path enumeration cost, resulting in overly long and difficult-to-manage test cases.

\subsubsection{Lack of Test Prioritization}
The generated test cases are not prioritized or ranked:
Although \TGen\/ provides comprehensive coverage, it does not account for execution order. 
Introducing prioritization
---
for example, executing main flows before exceptional paths
---
should improve testing efficiency and better align with real-world testing practices. 
Future work will explore extending CFGs with path-level priority to enable automated test-case ranking.

\subsubsection{Abstract Test Cases without Executable Scripts}
The approach generates abstract test cases to guide test engineers, but it does not include concrete input data or executable scripts. 
Linking abstract test cases to source code or runtime environments could enable the generation of fully executable test cases, thereby reducing manual effort and supporting end-to-end automation.

\section{Conclusions and Future Work
\label{sec:conclusion}}

This paper has proposed an end-to-end pipeline for generating test cases from NL use-case descriptions:
It is called \textit{Test Generation based on LLM-generated Control Flow Graphs} (\TGen\/).
\TGen\/ involves three steps:
(1) generating a CFG using an LLM; 
(2) enumerating feasible execution paths; and 
(3) creating test cases from the paths. 
Experiments on six use-case datasets demonstrated that LLMs can construct structurally-valid CFGs from NL requirements. 
Compared with direct LLM-based generation, and other existing pipelines, \TGen\/ generated more complete and structurally-coherent test cases, while better preserving execution flows and the requirements logic. 
Automated evaluations and practitioner studies both confirmed the practical effectiveness of the approach for RBTG.
Overall, the findings suggest that combining LLM-based semantic reasoning with verifiable intermediate representations provides an effective bridge between NL requirements and systematic test generation.

Our future work will include expanding support to a wider spectrum of datasets and diverse types of NL requirements, including user stories and unstructured requirement specifications. 
This will require enhancing the robustness of CFG generation to better handle noisy, incomplete, or domain-specific inputs.
Furthermore, incorporating interactive or adaptive mechanisms, such as human-in-the-loop feedback \cite{marques2024using}, may enable more comprehensive requirements analysis. 
Finally, establishing a unified LLM-driven framework that connects requirements analysis, model generation, and test creation, execution, and evaluation, is an interesting direction for self-adaptive and explainable test generation from requirements.
We look forward to engaging further with this goal.

\section*{Acknowledgment}
This work in this paper was partly supported by the Science and Technology Development Fund of Macau, Macao SAR (Grant Nos. 0069/2025/RIB2 and 0021/2023/RIA1), and the General Research Project of Zhejiang Provincial Department of Education (Grant No. Y202558065).

\bibliographystyle{ACM-Reference-Format}
\bibliography{LSCFG_TGen_ref}

\end{document}